\documentclass[aps, reprint, prx, superscriptaddress, 10pt,longbibliography]{revtex4-1}
\usepackage{graphicx}
\usepackage{epstopdf}
\usepackage{longtable} 
\usepackage{multirow}
\usepackage{hhline}
\usepackage{lipsum}
\usepackage{hyperref}
\usepackage{amsmath}
\usepackage{setspace}
\usepackage{xcolor}
\usepackage{ulem}
\usepackage[switch,columnwise]{lineno}
\begin{document}
\title{Signatures for  Berezinsky-Kosterlitz-Thouless  critical behaviour\\ in the planar antiferromagnet BaNi$_2$V$_2$O$_8$ }

\author{E. S. Klyushina}
\affiliation{Helmholtz-Zentrum Berlin f{\"u}r Materialien und Energie, 14109 Berlin, Germany} 
\affiliation{Institut f{\"u}r Festk{\"o}rperphysik, Technische Universit{\"a}t Berlin, 10623 Berlin, Germany} 
\author{J. Reuther}
\affiliation{Helmholtz-Zentrum Berlin f{\"u}r Materialien und Energie, 14109 Berlin, Germany} 
\affiliation{Dahlem Center for Complex Quantum Systems and Institut f{\"u}r Theoretische Physik, Freie Universit{\"a}t Berlin,
Arnimallee 14, 14195 Berlin, Germany}
\author{L. Weber}
\affiliation{Institute for Theoretical Solid State Physics, JARA-FIT, and JARA-HPC, RWTH Aachen University, 52056 Aachen, Germany}
\author{A.T.M.N. Islam}
\affiliation{Helmholtz-Zentrum Berlin f{\"u}r Materialien und Energie, 14109 Berlin, Germany} 
\author{J. S. Lord}
\affiliation{ISIS facility, STFC Rutherford Appleton Laboratory, Chilton, Didcot, Oxfordshire OX11 0QX, United Kingdom} 
\author{B. Klemke}
\affiliation{Helmholtz-Zentrum Berlin f{\"u}r Materialien und Energie, 14109 Berlin, Germany} 
\author{M. M{\aa}nsson}
\affiliation{Department of Applied Physics, KTH Royal Institute of Technology, SE-10691 Stockholm, Sweden} 
\author{S. Wessel}
\affiliation{Institute for Theoretical Solid State Physics, JARA-FIT, and JARA-HPC, RWTH Aachen University, 52056 Aachen, Germany}
\author{B. Lake}
\affiliation{Helmholtz-Zentrum Berlin f{\"u}r Materialien und Energie, 14109 Berlin, Germany} 
\affiliation{Institut f{\"u}r Festk{\"o}rperphysik, Technische Universit{\"a}t Berlin, 10623 Berlin, Germany} 

\begin{abstract}
We investigate the critical properties of the spin-$1$ honeycomb antiferromagnet BaNi$_2$V$_2$O$_8$, both below and above the ordering temperature $T_N$ using  neutron diffraction and muon spin rotation measurements. Our results characterize BaNi$_2$V$_2$O$_8$ as a two-dimensional (2D)  antiferromagnet across the entire temperature range, displaying a series of crossovers from 2D Ising-like to 2D XY and then to 2D Heisenberg behavior with increasing temperature. In particular, the extracted critical exponent of the order parameter reveals a narrow temperature regime close to $T_N$, in which the system behaves as a 2D XY antiferromagnet. Above $T_N$, evidence for Berezinsky-Kosterlitz-Thouless behavior driven by vortex excitations is obtained from the scaling of the correlation length. Our experimental results are in accord with classical and quantum Monte Carlo simulations performed for microscopic magnetic model Hamiltonians for BaNi$_2$V$_2$O$_8$. 
\end{abstract}

\maketitle

\section{Introduction} 

The Berezinsky-Kosterlitz-Thouless (BKT) transition is a paradigmatic example of a phase transition driven by topological defects. Due to its paramount importance in condensed matter physics, the underlying fundamental concepts of topology were recently distinguished by the Nobel prize in physics \cite{KosterlitzThouless,Kosterlitz}.
In low dimensional magnets, continuous spin rotation symmetry cannot be spontaneously broken at finite temperatures which, for example, rules out a finite-temperature transition to a conventional long-range ordered (LRO) state in a 2D Heisenberg magnet. While this famous result, known as the Mermin-Wagner theorem~\cite{Mermin}, crucially determines the role of low-energy fluctuations, it does not, however, apply to all types of phase transitions in low dimensions.  As predicted by Kosterlitz and Thouless and independently by Berezinsky, in a 2D  magnet with planar spins (such as in the classical XY model) a quasi-long-range ordered state with power-law  correlations exists below a finite transition temperature $T_\mathrm{BKT}$ \cite{KosterlitzThouless,BerezinskiiP1,BerezinskiiP2}. This thermal transition is driven by the proliferation and unbinding of  topological defects in the form of vortices. Below $T_\mathrm{BKT}$, these vortices are bound in vortex/antivortex pairs with opposite winding numbers. Above $T_\mathrm{BKT}$, these pairs deconfine into a plasma of mobile vortices, which manifest themselves through an exponential thermal decay of the correlation length $\xi(T)$~\cite{Kosterlitz}.

BKT phenomena were experimentally observed in physical realizations of the 2D XY model such as superfluids~\cite{Xu}, superconducting thin films~\cite{BKTthinFilms}, 2D organic magnetic complexes~\cite{BKTfield,opherden2020extremely} and more recently in a triangular lattice quantum Ising system \cite{BKTIsing}. However, no unambiguous solid-state prototype of the 2D XY model which develops BKT behavior has been established so far. This is typically due to the presence of additional terms in the Hamiltonian such as finite interplane coupling, which induces conventional three-dimensional (3D) magnetic LRO. Complications of this type are indeed encountered in various quasi-2D magnets such as K$_2$CuF$_4$, Rb$_2$CrCl$_4$, BaNi$_2$X$_2$O$_8$ (X = As, P) and MnPS$_3$ \cite{K2CuF4Hirakawa,Als_Nielsen_1993,Jongh,a13,Gaveau,RonnowMnPS3} in which BKT phenomena were intensively sought. Although these compounds are interesting candidate systems, their heat capacities have sharp $\lambda$-anomalies associated with a transition to 3D magnetic LRO~\cite{Yamada_Cv,BLOEMBERGEN,Regnault3}. Furthermore, they all exhibit 3D critical scaling and reveal a crossover to 3D XY or 2D XY regimes only {\it below} the transition temperature to conventional 3D LRO~\cite{K2CuF4crit,KLEEMANN,Als_Nielsen_1993,Jongh,Wildes2006}.

Even though interplane interactions are detrimental to BKT phenomena, Hakami {\it et\ al.}~\cite{Hikami} provided theoretical evidence that effective 2D XY behavior prevails over finite regions if these couplings are sufficiently small,  the size of these regions being related to the ratio of the intraplane to interplane couplings. Bramwell {\it et \ al.}~\cite{Bramwell}, studied a finite 2D XY magnet and showed that a transition to spontaneous finite magnetization occurs in the absence of interplane coupling,  characterized by an effective 2D XY exponent of $\beta=0.23$. This transition occurs above the bulk $T_\mathrm{BKT}$. Other aspects of the infinite 2D XY system, such as the presence of vortices and the characteristic scaling of the correlation length are not effected by finite-size effects. Thus, a  real magnetic compound may be used to explore BKT physics if the interplane coupling is sufficiently weak to allow 2D behavior over large length scales, e.g., comparable to the magnetic domain size.

Apart from interplane couplings, the realization of intralayer interactions of purely XY-type could pose another obstacle for the observation of BKT behavior in real solid state materials. However, recent quantum Monte Carlo simulations reveal that BKT behavior is still present in 2D Heisenberg magnets perturbed by an easy-plane anisotropy, even if this anisotropy is very weak~\cite{Cuccoli}. Hence, approximate 2D XY magnets, affected by a combination of small interplane couplings, finite size magnetic domains and weak easy-plane anisotropies can still display BKT phenomena even if they eventually develop magnetic LRO at low temperatures.

Here, we report a comprehensive investigation of the spin-1 honeycomb compound BaNi$_2$V$_2$O$_8$, which was recently discovered to be a rare physical realization of the 2D Heisenberg antiferromagnet (AFM) with XY exchange anisotropy and negligible interlayer coupling~\cite{Rogado,KlyushinaBaNi}. Using a combination of experimental and theoretical techniques we (i) establish a consistent phase diagram of BaNi$_2$V$_2$O$_8$, (ii) identify the temperature range over which it behaves as a 2D XY magnet and (iii) provide signatures of BKT scaling behavior driven by vortices in BaNi$_2$V$_2$O$_8$ within a 2D XY regime above $T_N$. 

BaNi$_2$V$_2$O$_8$ has trigonal crystal structure (space group R$\bar{3}$), where the $S=1$ Ni$^{2+}$ magnetic ions form honeycomb layers that are stacked perpendicular to the c-axis. The Hamiltonian was shown to have strong AFM 1$^\mathrm{st}$-neighbor ($J_\mathrm{n}=12.3$~meV), weaker AFM 2$^\mathrm{nd}$-neighbor ($J_\mathrm{nn}=1.25$~meV) and very weak 3$^\mathrm{rd}$-neighbor ($J_\mathrm{nnn}=0.2$~meV) intralayer Heisenberg couplings. The interlayer coupling if present is extremely weak with an upper limit on its magnitude of  $|J_\mathrm{out}|< 10^{-4} J_\mathrm{n}$ \cite{KlyushinaBaNi}. In addition, a weak single-ion  XY-anisotropy ($D_\mathrm{EP(XY)}=0.0695$~meV) favours spin directions within the honeycomb plane, while an even weaker easy-axis anisotropy  ($D_\mathrm{EA}=-0.0009$~meV) selects three equivalent in-plane directions~\cite{KlyushinaBaNi}. The system develops conventional N\'eel long-range magnetic order first reported below $T_N=50$~K based on powder neutron diffraction \cite{Rogado}. Here, we identify $T_N=47.75\pm0.25$~K from single crystal neutron diffraction (Appendix~\ref{sec:TNneutrons}) and muon spin rotation measurements (Appendix~\ref{sec:TNmuons}). 

Further indications for the 2D Heisenberg behavior are provided by the heat capacity of BaNi$_2$V$_2$O$_8$, which does not display sharp features at $T_N$ \cite{Rogado,Specificheat2D}. Moreover, recent single crystal static magnetic susceptibility measurements reveal planar anisotropic magnetic behaviour above $T_N$, suggesting that BaNi$_2$V$_2$O$_8$ is a promising candidate to realize the 2D Heisenberg model with XY anisotropy at finite temperatures and, therefore, could host BKT physics~\cite{KlyushinaBaNi}. Thus far, the relevance of the BKT scenario was experimentally explored using electron spin resonance and nuclear magnetic resonance measurements,  reporting values of  $T_\mathrm{BKT} = 43.3$~K \cite{Heinrich} and $40.2\pm0.5$~K \cite{Waibel2015}, respectively. 
On the other hand, the magnetic properties of BaNi$_2$V$_2$O$_8$ at finite temperatures,  such as  the order parameter  and  correlation length scaling, have not been studied so far. Here we report on a comprehensive experimental investigation using neutron scattering and susceptibility measurements, which demonstrate that BaNi$_2$V$_2$O$_8$ is a rare example of a 2D AFM at all temperatures. We also performed classical (CMC) and quantum Monte Carlo (QMC) simulations, which are in accord with the experimental observations and provide further support for the BKT scenario.

\section{Methods} 

Single crystals of BaNi$_2$V$_2$O$_8$ were grown in the Core Lab for Quantum Materials (QMCL) at the Helmholtz-Zentrum Berlin f{\"u}r Materialien und Energie. Zero-field (ZF) muon spin rotation ($\mu^{+}SR$) measurements were performed on a single-crystal sample  using the EMU $\mu^{+}SR$ spectrometer at the ISIS Neutron and Muon Source, UK. The sample was oriented so that the muon beam was perpendicular to the honeycomb plane and the muon spectra were measured over the temperature range 8 - 48.5~K (Appendix~\ref{sec:ZFmuons}). Weak transverse field (TF) $\mu^{+}SR$ measurements were also performed over 45 - 100~K (Appendix~\ref{sec:TNmuons}). Elastic neutron scattering measurements were performed over the temperature range 1.47 - 56~K on the cold neutron triple-axis spectrometer TASP at the Paul Scherrer Institute (PSI), Switzerland \cite{TASP} (Appendix~\ref{sec:Methods}). The correlation length was also explored in the range of  48 - 140~K using the TASP in  two-axis mode (Appendix~\ref{sec:Methods}). The static magnetic susceptibility was measured at the QMCL, over the range 2 - 640~K as discussed in Ref.~\cite{KlyushinaBaNi}. For comparison, CMC (Appendix~\ref{sec:CMC}) and QMC (Appendix~\ref{sec:QMC}) simulations were performed, based on model Hamiltonians of BaNi$_2$V$_2$O$_8$.

\begin{figure*}
\includegraphics[width=1\linewidth]{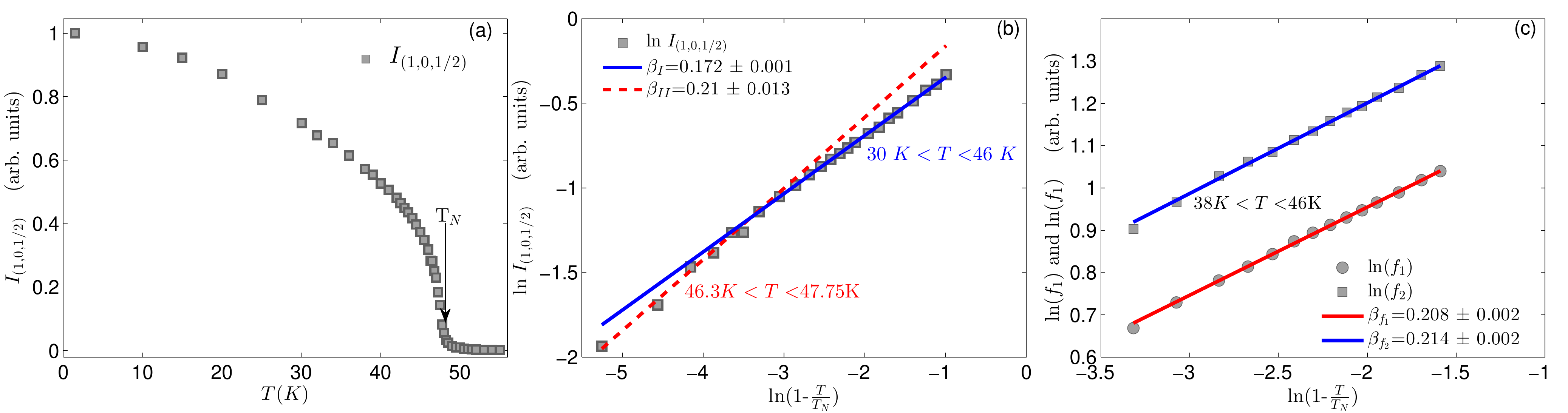}
\caption{Integrated intensity $I_{(1,0,1/2)}$ of the (1,0,1/2) magnetic Bragg peak measured by neutron scattering plotted (a) as a function of temperature and (b) as a function of the reduced temperature in a logarithmic scale. (c) The logarithm of the frequencies $f_1$ and $f_2$ extracted from the ZF-$\mu^+$SR spectra and plotted as functions of the reduced temperature in a logarithmic scale.}
\label{fig:Fig1} 
\end{figure*}

\section{results}
We first report the results of the neutron scattering and muon spin rotation measurements, separately below and above the magnetic transition temperature $T_N$. This is  followed by a detailed comparison to the theoretical results from  CMC and QMC simulations. 

\subsection{Magnetic scaling below $T_N$} 

We first examine the magnetic properties of BaNi$_2$V$_2$O$_8$ below  $T_N$ (Appendix~\ref{sec:Methods}). Figure~\ref{fig:Fig1}(a) shows the integrated intensity  $I_{(1,0,\frac{1}{2})}$  of the (1,0,$\frac{1}{2}$) magnetic Bragg peak,  as a function of temperature $T$. The intensity smoothly decreases with increasing temperature, and starts to drop steeply near 47~K. Above $T_N$, some residual intensity remains that decreases gradually to zero, reminiscent of the behaviour predicted for finite-size 2D XY magnets (see Fig.~1 in Ref.~\cite{Bramwell}).
Note that this signal is not critical scattering because these measurements were performed with an analyser. Figure~\ref{fig:Fig1}(b) shows $I_{(1,0,\frac{1}{2})}$ on a logarithmic scale as a function of the reduced temperature, $t = (T-T_N)/{T_N}$. For single power law behavior $I_{(1,0,\frac{1}{2})} \propto |t|^{2\beta}$ with the critical exponent $\beta$, the logarithm would follow a linear dependence, $\ln I_{(1,0,\frac{1}{2})}\propto 2\beta \ln |t|$, whose slope is set by the value of  $\beta$.
However, we observe that  $\ln I_{(1,0,\frac{1}{2})}$ does not follow a single straight line but reveals a crossover around 46 - 46.3~K, which separates two temperature regions, (I) 30 - 46~K, and (II) 46.3 - 47.5~K. Within both regimes, $\ln I_{(1,0,\frac{1}{2})}$ can be fitted independently to a linear $\ln |t|$-dependence, with effective critical exponents, $\beta_\mathrm{I}$=0.172$\pm$0.001 and $\beta_\mathrm{II}$=0.21$\pm$0.013, respectively. Remarkably, just below $T_N$, the critical exponent $\beta_\mathrm{II}$ is thus close to the value $\beta_\mathrm{XY}$=0.23 predicted for a large but finite 2D XY system \cite{Bramwell}. In a real material like BaNi$_2$V$_2$O$_8$ such finite-sized effects could arise from the formation of domains.
In contrast, the critical exponent $\beta_{I}$ which characterizes the temperature region below $T_\mathrm{EA}=46$~K,  resides between the value for the 2D XY ($\beta$=0.23) and the 2D Ising model ($\beta$=0.125). We attribute this tendency towards the 2D Ising exponent to the presence of a weak in-plane easy-axis anisotropy in BaNi$_2$V$_2$O$_8$~\cite{KlyushinaBaNi}. Also note that a recent theoretical study indeed predicts a continuous range of critical exponents $0.125<\beta<0.23$ for XY magnets with in-plane easy-axis anisotropy \cite{Taroni}.

The critical properties of  BaNi$_2$V$_2$O$_8$ were further investigated by analysing ZF-$\mu^+$SR spectra of BaNi$_2$V$_2$O$_8$  over the temperature range 38 - 46~K (the spectra above 46~K were found to be unreliable). Two distinct frequencies were identified in the muon spectrum, which can be attributed to the presence of two muon stopping sites, i.e.,  the muons experience two distinct internal magnetic fields, which both  directly scale with the long-range magnetic order  (Appendix~\ref{sec:ZFmuons}). 

Figure~\ref{fig:Fig1}(c) shows the temperature dependence of both  frequencies $f_1$ and $f_2$  on a logarithmic scale as functions of $\ln|t|$. The best power-law fits were achieved for the slopes 
$\beta(f_1)=0.208 \pm 0.002$ and $\beta(f_2)=0.214 \pm 0.002$, respectively. These muon results suggest that the 2D XY regime in BaNi$_2$V$_2$O$_8$ persists down to 38~K, in contrast to the neutron data, which suggest a tendency toward Ising-like behavior below $T_\mathrm{EA}=46$~K.  
This difference can be attributed to the different time-scales probed by muon and neutron spectroscopy: The neutrons are faster than the muons and therefore slow fluctuations appear effectively static for the neutrons, while the muons which are more sensitive, correctly identify them as dynamic. Indeed, a comparison of the neutron and muon data reveals that the muons observe a lower magnetization than the neutrons (Appendix~\ref{sec:Magnetization}), further supporting this point. We can imagine a scenario in which just below $T_{N}$ the spins order antiferromagnetically along a general direction in the XY plane with fluctuations about this direction, while below $T_{EA}=46$~K the fluctuations occur predominantly toward the easy-axis directions, giving rise to  the reduction of the critical exponent observed in the neutron data. This quenching however, does not affect the critical exponents extracted from the ZF-$\mu^+$SR spectra, as the muons assign these fluctuations to spin dynamics. 

The value of the scaling exponent of approximately 0.21 near $T_N$, extracted from both the neutron and muon data, falls slightly below  the value of 0.23  quoted for a 2D XY finite-sized system. It should be mentioned however, that this effective exponent actually varies within the range 0.21 - 0.23, depending on the size of the finite system, and a value of  0.23 would be expected only for larger domains~\cite{Bramwell}.
As  will be demonstrated  below, a 2D XY scaling regime also emerges for temperatures just above $T_N$, further suggesting the relevance of BKT physics for BaNi$_2$V$_2$O$_8$.

\subsection{Magnetic scaling above $T_N$} 

\begin{figure}
\includegraphics[width=1\linewidth]{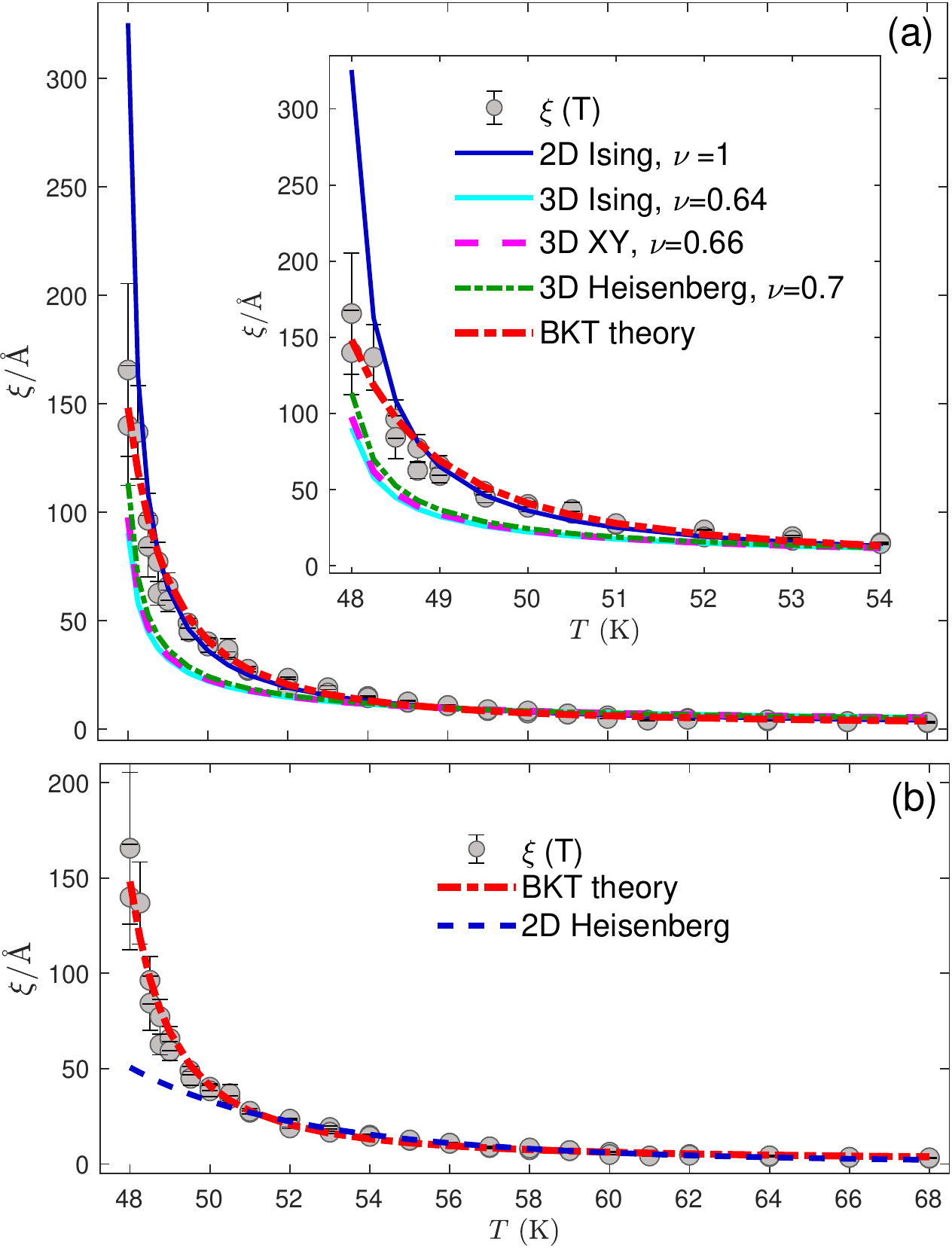}
\caption{Correlation length $\xi(T)$  as a function of temperature $T$. (a) compares the fit of the BKT expression (goodness of fit, $\chi^{2}=7.5$) to fits of conventional power law scaling with an exponent fixed to 
$\nu=1$ (2D Ising, $\chi^{2}=13.2$),
$\nu=0.64$ (3D Ising, $\chi^{2}=62.7$),  
$\nu=0.66$ (3D XY, $\chi^{2}=58.7$), and 
$\nu=0.7$ (3D Heisenberg, $\chi^{2}=50.2$).
The inset shows the low temperature region in detail.
(b) compares the fits to the 2D Heisenberg model ($\chi^{2}=9.7$) and  the BKT expression ($\chi^{2}=7.5$).}
\label{fig:FigCorrLength} 
\end{figure}

To quantify the magnetic properties of BaNi$_2$V$_2$O$_8$ above $T_N$, the thermal decay of the spin-spin correlations was investigated. The correlation length $\xi(T)$ was extracted as the inverse full-width-at-half-maximum (FWHM) of the energy-integrated magnetic signal at wavevector ${(1,0,\frac{1}{2})}$, measured over the temperature range 48 - 140~K. In the following, we compare $\xi(T)$  to various theoretical scaling forms.  However, such  theoretical expressions for $\xi(T)$ are typically based on  continuum descriptions and as such, apply when $\xi(T)$ extends
 well beyond the microscopic lattice scale, which for BaNi$_2$V$_2$O$_8$ is set by the shortest distance $d_\mathrm{Ni} = 2.90$~\AA\  between neighboring Ni$^{2+}$ ions within the {\it{ab}}-plane.
Therefore, in the following, we consider $\xi(T)$ only over the temperature range 48 - 68~K, where the condition $\xi > d_{\mathrm{Ni}}$ is satisfied (the gray filled circles in Fig. \ref{fig:FigCorrLength}). The correlation length over the entire temperature range from 48 to 140~K  is provided in  Appendix~\ref{sec:CorrelationLength}. 

Near criticality, the correlation length typically follows a  power law scaling $\xi \propto t^{\nu}$ as a function of the reduced temperature $t=({T-T_N})/{T_N}$. Here, the correlation length exponent $\nu$   characterizes the universality class of the thermal phase transition. In particular, $\nu$ takes on  the value $\nu=1, 0.64, 0.66, 0.7$ for the 2D Ising, 3D Ising, 3D XY  and 3D Heisenberg  universality class, respectively \cite{collins}. We first observed that such a single power law scaling is inappropriate to describe the thermal decay of the correlation length in BaNi$_2$V$_2$O$_8$ within the considered temperature range of 48 - 68~K. As shown in Fig.~\ref{fig:FigCorrLength}, the fits to 3D Ising, 3D XY and as well as 3D Heisenberg scaling deviate significantly from the data below 54~K. In contrast, the 2D Ising scaling clearly overestimates the data at the lower temperatures, although it does yield better agreement than the other power laws. A similar analysis of the correlation length on a logarithmic scale as a function of $\ln t $ confirms that $\xi(T)$ is not fitted well by any single power law within the relevant temperature regime 48 - 68~K (Appendix~\ref{sec:AlgebraicScaling}). 

As a next step, $\xi(T)$ was fitted to the expression for the 2D Heisenberg magnet inside the classical regime~\cite{2DHcor}:
\begin{equation}
{\xi(T) \sim \exp\Bigg(\frac{2\pi\rho_s}{k_BT}\Bigg)\cdot\Bigg(1-\frac{k_BT}{4\pi\rho_s}+O(T)^2\Bigg)}.
\label{eq:1}
\end{equation}\
Here, $\rho_s$ is a non-universal number,  quantifying an effective spin-stiffness. 
The dashed blue line through the data in Fig.~\ref{fig:FigCorrLength} shows the best fit of $\xi(T)$ to the above expression, achieved for  
$\rho_s = 7.01 \pm 0.23$~meV  over the range 48 - 68~K . These results imply that the 2D Heisenberg model gives a good description of $\xi(T)$ for temperature above  51 - 52~K, even though it does not take into account the  anisotropies and  the interlayer coupling. However,  this isotropic model  does not reproduce the experimental data in the lower temperature regime closer to $T_{N}$, an observation that can be attributed to the planar anisotropy. 

Since neither a conventional power law nor the 2D Heisenberg model scaling describe the spin-spin correlations of BaNi$_2$V$_2$O$_8$ accurately over the temperature range just above the $T_N$, we also analyzed $\xi(T)$  in terms of the BKT exponential scaling law for the 2D XY model, which reads
\begin{equation}
 \xi(T) \sim \: \exp\Bigg(b\:\sqrt{\frac{T_\mathrm{BKT}}{T-T_\mathrm{BKT}}}\Bigg).
\label{eq:2}
\end{equation}
Here, $b$ is a  non-universal number and  $T_{BKT}$ the BKT transition temperature. The dashed-dotted red line in Fig.~\ref{fig:FigCorrLength} presents the best fit of Eq.~\eqref{eq:2} to 
the experimental data for BaNi$_2$V$_2$O$_8$, achieved with $T_\mathrm{BKT}=44.95 \pm 0.11$~K, and $b$ set to 1.5~\cite{Kosterlitz}, respectively. We indeed find that   the BKT scaling  of $\xi(T)$ accurately follows the thermal decay of $\xi$ over the entire explored temperature range, 48  - 68~K. A comparison of the BKT model expression to the other scenarios thus reveals that it describes the magnetic fluctuations of BaNi$_2$V$_2$O$_8$ significantly better than the 2D Heisenberg model or a conventional power law. The superiority of BKT model over the power laws is further confirmed by the analysis of $\xi(T)$ on a logarithmic scale given in Appendix~\ref{sec:BKTScaling1}. 
     
Finally, $\xi(T)$ was fitted to the  BKT expression over several temperature ranges extending from 48~K up to $T_\mathrm{max}$, using different values of $T_\mathrm{max}$=55, 60, 66~K, in order to assess the robustness of the extracted value of $T_\mathrm{BKT}$. These fits are provided in Appendix~\ref{sec:BKTScaling2} and reveal that $T_\mathrm{BKT}$ lies within the range 44.44~K$<T_\mathrm{BKT}<$44.95~K, where $T_\mathrm{BKT}=44.44$~K and  $T_\mathrm{BKT}=44.95$~K are extracted for the temperature ranges 48 - 55~K and 48 - 68~K, respectively. We  take the mean value of $T_\mathrm{BKT}=44.70 \pm0.25$~K  as our best estimate for the BKT transition temperature. 
Since $T_\mathrm{BKT}$ is lower than $T_N$, the quasi-ordered state is in fact hidden by the onset of LRO at $T_N$. Nevertheless, deconfined vortex/anti-vortex excitations are expected to occur in the regime just above $T_N$, which we indeed quantify below using a microscopic  model description for the magnetism in BaNi$_2$V$_2$O$_8$.

\subsection{Comparison with microscopic models}

To further benchmark the BKT physics in BaNi$_2$V$_2$O$_8$ with respect to microscopic details, classical (CMC) and quantum (QMC) simulations were performed, based on model Hamiltonians for BaNi$_2$V$_2$O$_8$ in order to (i) compare with the magnetic susceptibility recently measured on a single crystal~\cite{KlyushinaBaNi} and (ii) verify the values of $T_\mathrm{BKT}$ extracted from the analysis of $\xi(T)$.

\begin{figure}
\includegraphics[width=1\linewidth]{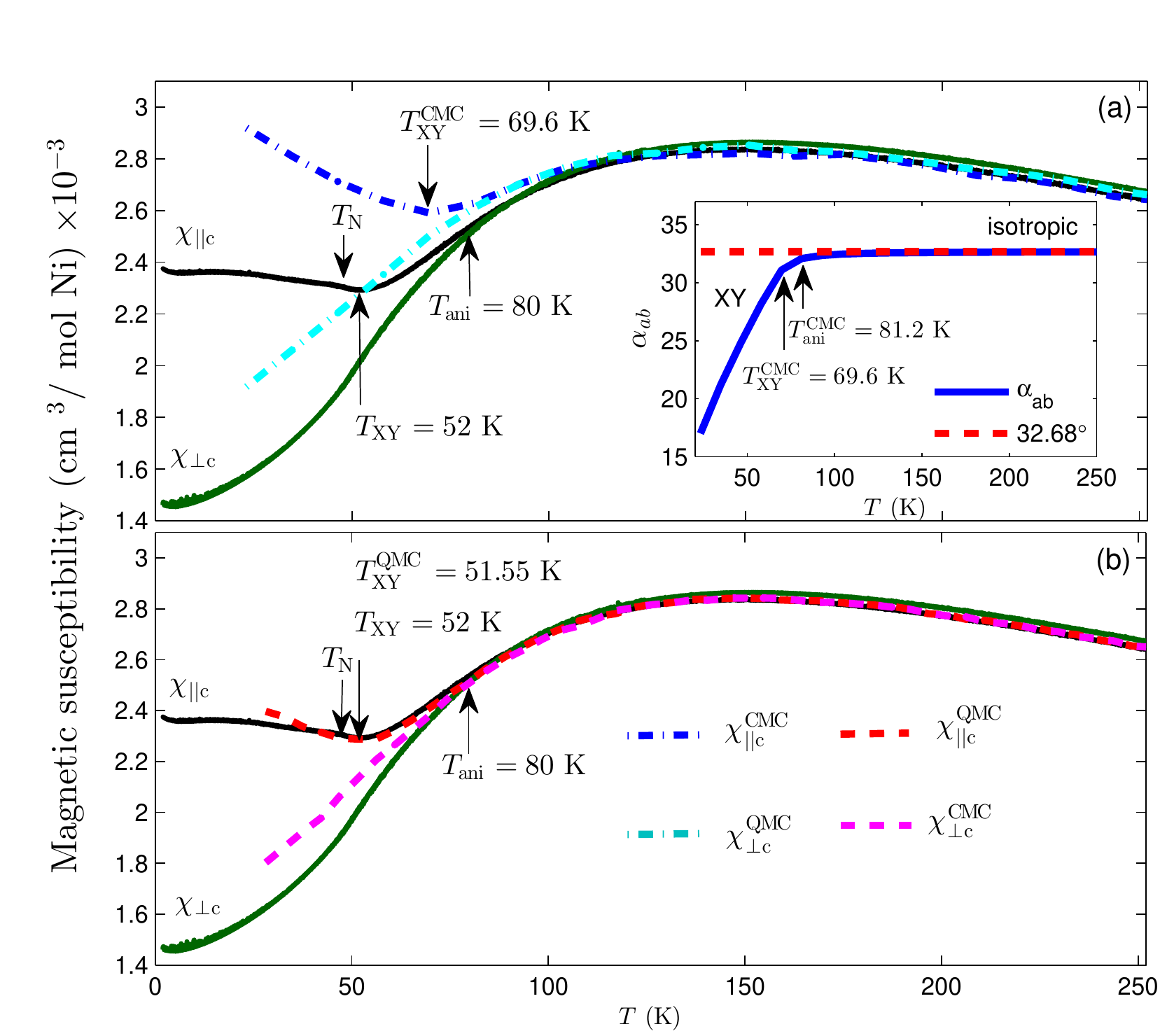}
\caption{The magnetic susceptibility of BaNi$_2$V$_2$O$_8$ measured in a magnetic field of 1~T applied parallel (solid black line) and perpendicular (solid green line) to the c-axis \cite{KlyushinaBaNi}. The dashed-dotted blue and cyan lines (dashed red and magenta lines) on panel (a) [(b)] show the results of CMC (QMC) computations, respectively. The inset shows the average angle $\alpha_{ab}$ between the magnetic moments and the honeycomb plane as computed using CMC.} 
\label{fig:Fig2} 
\end{figure}

Figures~\ref{fig:Fig2}(a) and~(b) show comparisons of the CMC and QMC results to the experimental data, respectively. The solid black and green lines present the magnetic susceptibility for a constant magnetic field of $B=1$~T, applied parallel ($\chi_{||c}$) or perpendicular ($\chi_{\perp c}$) to the c-axis, respectively,  where the c-axis is perpendicular to the easy plane~\cite{KlyushinaBaNi}.  At high temperature, the susceptibility of BaNi$_2$V$_2$O$_8$  behaves isotropically and the broad maximum at 150~K is attributed to low-dimensional spin-spin correlations. Upon decreasing $T$ below $T_\mathrm{ani} \approx 80$~K, the susceptibilities $\chi_{||c}$ and $\chi_{\perp c}$ split, revealing that the planar anisotropy is already evident well above $T_N$. The out-of-plane susceptibility $\chi_{||c}$ has a minimum at $T_\mathrm{XY}=52$~K, which is attributed to the crossover to a regime dominated by the XY-anisotropy, below which the spins lie mostly within the honeycomb easy-plane, according to recent QMC simulations performed for the $S=\frac{1}{2}$ square lattice~\cite{Cuccoli}. Indeed, this is consistent with the previous section where we also found that below about 51~K, the isotropic 2D Heisenberg model scaling fails to follow the correlation length $\xi(T)$ in BaNi$_2$V$_2$O$_8$.

The dashed-dotted blue and cyan lines in Fig.~\ref{fig:Fig2}(a) present the CMC results for $\chi_{||c}$ and $\chi_{\perp c}$, respectively, using the Hamiltonian for BaNi$_2$V$_2$O$_8$, but without the interlayer coupling (Appendix~\ref{sec:CMC}). Both $\chi^\mathrm{CMC}_{\perp c}$ and $\chi^\mathrm{CMC}_{||c}$ are in good agreement with the experimental data at high temperatures. In particular, the position of the broad maximum matches the experimental value very well. Below this maximum, $\chi^\mathrm{CMC}_{||c}$ and $\chi^\mathrm{CMC}_{\perp c}$ decrease smoothly, revealing an anisotropic splitting around $T_\mathrm{ani}\approx 80$~K, as found also in the experimental data. This characteristic temperature can be quantified from the computed average angle between the spins and the easy-plane. At high temperature this angle $\alpha_{ab}$ resides at $32.7^\circ$, corresponding to the average out-of-plane component for a randomly oriented three-component spin [cf. the inset of Fig.~\ref{fig:Fig2}(a)]. Below $T_\mathrm{ani}^\mathrm{CMC} \approx 80$~K,
$\alpha_{ab}$  starts to decrease, clearly indicating the onset of the easy-plane behavior. At lower temperatures, only qualitative agreement is observed between the experimental data and the CMC calculations. Indeed, $\chi^\mathrm{CMC}_{||c}$ displays the characteristic minimum at $T^\mathrm{CMC}_\mathrm{XY}\approx 70$~K which is somewhat higher than the experimental value of $T_\mathrm{XY}=52$~K. This difference is attributed to the neglect of quantum fluctuations in the CMC simulations. 

\begin{figure}
\includegraphics[width=1\linewidth]{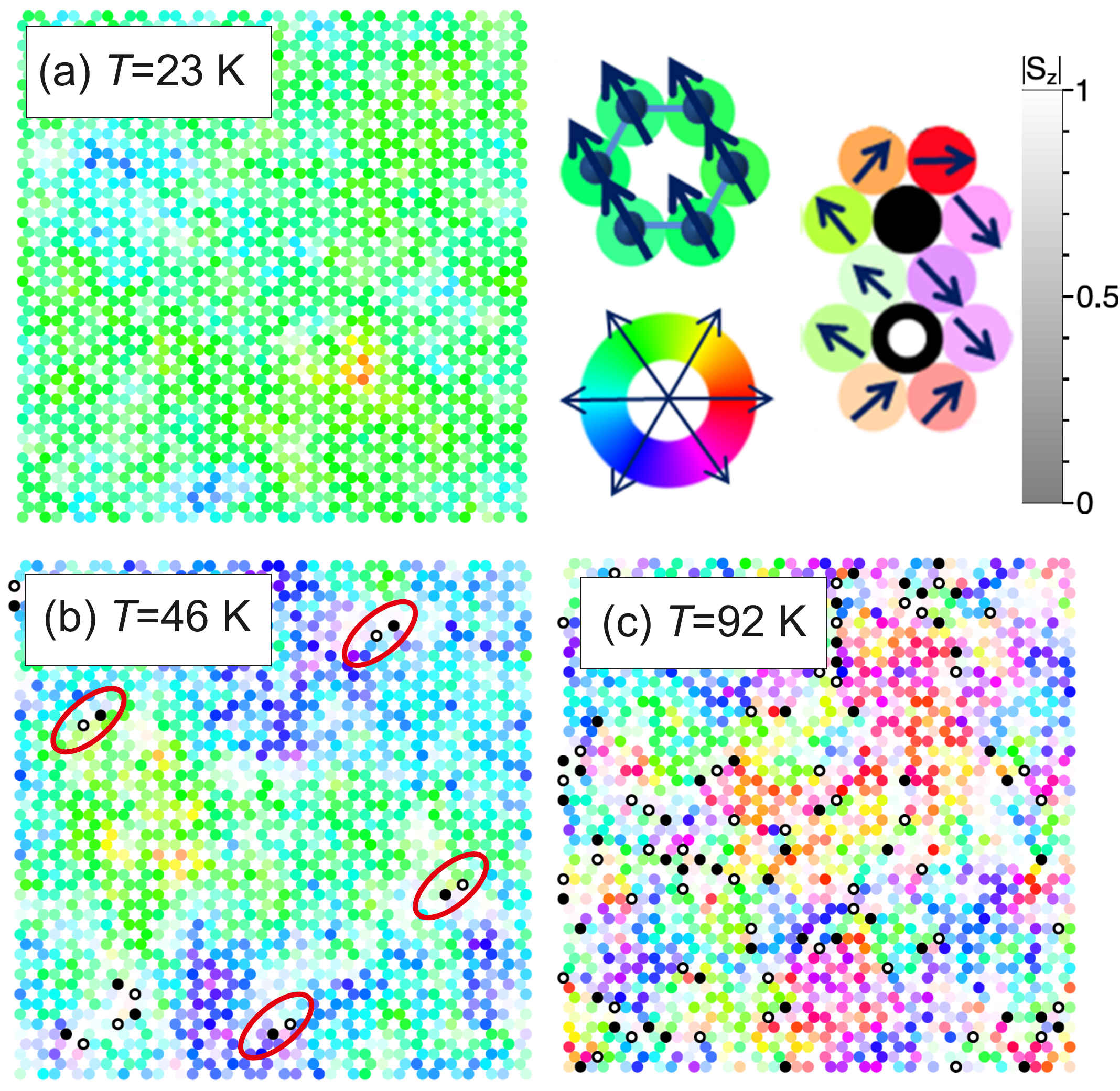} 
\caption{(a)-(c) Example configurations from CMC simulations of the honeycomb lattice where every second spin is artificially flipped for simplicity at (a) $T = 23$~K,  (b) $T = 46$~K and (c) $T = 92$~K. The spin directions are indicated by colors, and  the intensity of the color quanifies the size of the out-of-plane component. Closed (open) black circles indicate  vortices (antivortices) and the red ring highlight  vortex-antivortex pairs.} 
\label{fig:Fig3} 
\end{figure}

Before quantifying further the effects of quantum fluctuation in terms of QMC simulations, we demonstrate that 
the CMC computations support the presence of spin-vortex states in BaNi$_2$V$_2$O$_8$ at finite temperatures. Figures~\ref{fig:Fig3}(a), (b), and (c) show example CMC real-space configurations for $T = 23, 46$ and 92~K, respectively. The CMC simulations reveal a conventionally ordered AFM ground state at $T=0$~K. According to the BKT theory, the density of spin-vortex excitations is low at small temperatures and, indeed, we observe no vortices within the computed domain at $T=23$~K. Upon increasing temperature, a  finite density of bound vortex-antivortex pairs is observed at $T=46$~K. For $T=92$~K, the density of vortex excitations is significantly larger and they now form a deconfined plasma, i.e., their  binding into vortex-antivortex pairs is no longer discernible. These observations clearly reveal that BKT physics is relevant for the magnetism of BaNi$_2$V$_2$O$_8$. 

We  estimate the BKT transition temperature within the CMC simulations based on the real space spin-spin correlation function $C(r)$. BKT theory predicts that $C(r)$ decays below $T_\mathrm{BKT}$ as a function of spin separation $r$ according to a power law $C(r)\propto r^{-\eta (T)}$, where $\eta=1/4$ at $T_\mathrm{BKT}$. Based on this criterion, we get $T^\mathrm{CMC}_\mathrm{BKT}=55$~K. This temperature is again higher than $T_\mathrm{BKT}= 44.695 \pm 0.255$ K estimated from fitting the experimental $\xi(T)$ above $T_{N}$. This difference can again be attributed to the fact that CMC does not account for quantum fluctuations, which we would expect to reduce $T_\mathrm{BKT}$.

To account for the presence of quantum fluctuations in our theoretical modeling of the magnetism in  BaNi$_2$V$_2$O$_8$, QMC simulations were performed for  this $S=1$ system (Appendix~\ref{sec:QMC}). However, in order to avoid the sign problem in QMC, a simplified Hamiltonian for BaNi$_2$V$_2$O$_8$ has to be used, which  includes  only the 1$^\mathrm{st}$-neighbor interaction $J_\mathrm{n}$ and the easy-plane anisotropy $D_\mathrm{EP(XY)}$. The dashed red and magenta lines in Fig.~\ref{fig:Fig2}(b) present the QMC simulations of the magnetic susceptibility parallel ($\chi^\mathrm{QMC}_{||c}$) and perpendicular ($\chi^\mathrm{QMC}_{\perp c}$) to the c-axis, respectively. 
The best agreement with the experimental data was achieved for $J^\mathrm{QMC}_\mathrm{n}=8.07$~meV and $D^\mathrm{QMC}_\mathrm{EP(XY)}=0.04556$~meV (Appendix~\ref{sec:QMCvsData}).  Note that $J^\mathrm{QMC}_\mathrm{n}$ is significantly smaller than the coupling $J_\mathrm{n}=12.3$~meV of the original Hamiltonian for BaNi$_2$V$_2$O$_8$. This difference can be attributed to the exclusion of the frustrated interactions $J_\mathrm{nn}$ as well as $J_\mathrm{nnn}$. Indeed, when the spin-wave dispersions of BaNi$_2$V$_2$O$_8$ are fitted using the simplified Hamiltonian, the best fit is achieved for $J_\mathrm{n}=8.8$~meV and $D_\mathrm{EP(XY)}=0.099$~meV, in good agreement with the QMC estimates (Appendix~\ref{sec:SimpHam}). 
 
A comparison of the experimental data with the scaled susceptibilities $\chi^\mathrm{QMC}_{\perp c}$ and $\chi^\mathrm{QMC}_{||c}$ reveals remarkably good quantitative agreement over the full temperature range. $\chi^\mathrm{QMC}_{\perp c}$ and $\chi^\mathrm{QMC}_{||c}$ display an anisotropic splitting that matches the one observed in the experimental data below $T_\mathrm{ani}\approx 80$~K. Furthermore, $\chi^\mathrm{QMC}_{||c}$ shows the characteristic minimum at $T^\mathrm{QMC}_\mathrm{XY}=51.55$~K, which is in accord with the experimental value $T_\mathrm{XY}=52$~K. The nature of this minimum was verified by performing QMC computations for the Hamiltonian without the $D_\mathrm{EP(XY)}$ term. The results shown in  Appendix~\ref{sec:QMCEasyPlane} reveal no minimum in the out-of-plane susceptibility $\chi^\mathrm{QMC}_{||c}$ for $D_\mathrm{EP(XY)}=0$, hence confirming the connection between the minimum at $T_\mathrm{XY}$ and the crossover to the planar regime. Based on the  spin-spin correlation function $C(r)$ from the QMC simulations, we furthermore extract $T^\mathrm{QMC}_\mathrm{BKT}=40.2$~K, which is in reasonable agreement with $T_\mathrm{BKT} =44.70 \pm 0.25$~K extracted from the experimental data.   

\section{Discussion}

\begin{figure}
\includegraphics[width=1\linewidth]{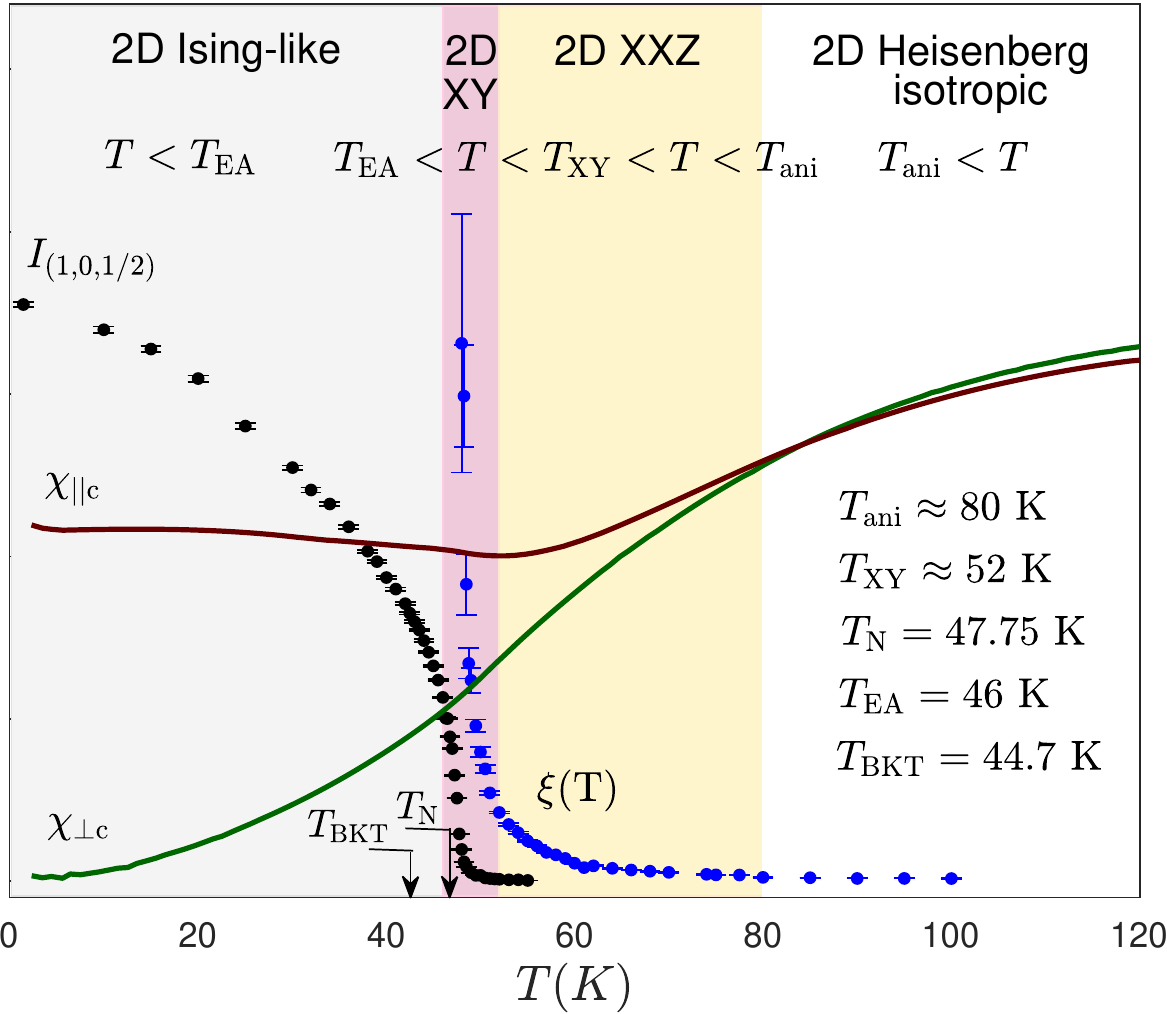} 
\caption{Phase diagram of BaNi$_2$V$_2$O$_8$ as obtained from our experiments where the different phases are identified by the different shaded colors. Filled black and blue circles show the temperature dependence of the integrated intensity $I_{(1,0,1/2)}$ of the (1,0,1/2) magnetic Bragg peak and the correlation length, respectively. The solid dark red and green lines show the magnetic susceptibility measured in applied fields perpendicular and parallel to the c-axis, respectively.} 
\label{fig:Fig4} 
\end{figure}

 Our experimental investigation reveals that BaNi$_2$V$_2$O$_8$ behaves as an ideal 2D magnet over the explored temperature range up to 140~K. A corresponding phase diagram as a function of temperature is presented in Fig.~\ref{fig:Fig4}, displaying several distinct temperature regimes, in which  various anisotropies become relevant. The correlation length in combination with magnetic susceptibility, and supported by the results of classical and quantum Monte Carlo simulations, reveal that BaNi$_2$V$_2$O$_8$ behaves as an isotropic 2D Heisenberg magnet at high temperatures above $T_\mathrm{ani}\approx 80$~K. A weak XY-anisotropy is observable below $T_\mathrm{ani}$ which becomes significant below $T_\mathrm{XY}=52$~K defining thus a 2D XXZ regime with only a weak planar anisotropy for $T_\mathrm{XY}<T<T_\mathrm{ani}$, and a 2D XY regime with dominant planar fluctuations for $T<T_\mathrm{XY}$. The critical exponent of the order parameter, as extracted from the elastic neutron measurements, reveals that the 2D XY behavior extends below $T_N=47.75$~K down to $T_\mathrm{EA}=46$~K, defining thus  a 2D XY regime  for $T_\mathrm{EA}<T<T_\mathrm{XY}$. Below $T_\mathrm{EA}$, the effective exponent $\beta$ from neutron measurements tends towards the theoretical value of the 2D Ising model, which can be associated with the presence of Ising-like fluctuations due to the weak in-plane easy-axis anisotropy $D_\mathrm{EA}$. This signature of  2D Ising-like behaviour is however not observed in the muon measurements which instead suggest that the 2D XY regime extends down to much lower temperatures. This disagreement is attributed to the different time scale of the muon and neutron probes and indicates that below $T_\mathrm{EA}=46$~K the magnetic moments are actually slowly fluctuating towards the easy-axis directions, rather than statically pointing along them. 
 It is worth emphasizing that due to its six-fold symmetry the in-plane easy-axis anisotropy is not expected to suppress the BKT behavior~\cite{Jose}.

We now discuss the nature of the phase transition to static magnetic order at $T_N$ in BaNi$_2$V$_2$O$_8$, which is characterized by the 2D XY critical exponent and, therefore, is not induced by the 2D Ising easy-axis anisotropy or 3D couplings. Such a transition is prohibited by the 
Mermin-Wagner theorem in the thermodynamic limit of an ideal 2D XY magnet, however, as shown by  Bramwell {\it et.\ al.}~\cite{Bramwell}, spontaneous static magnetization always occurs in a finite system, even in the absence of interplane coupling. These finite regions can be large and in a real material like BaNi$_2$V$_2$O$_8$ could be due to domains. The domains might either be static, such as structural domains, or reflect the existence of more dynamic and temperature dependent magnetic domains. In the presence of a weak interplane coupling $J_\mathrm{out}$, the domain lengthscale $L_\text{domain}$, must be smaller than $L_\mathrm{eff}=d_\mathrm{Ni}\sqrt{J_\mathrm{n}/|J_\mathrm{out}|}$ where $J_\mathrm{n}$ is the intraplane coupling, 
in order for the transition to retain its 2D XY character. The relevance of this scenario for BaNi$_2$V$_2$O$_8$ is suggested by the agreement of the measured critical exponent $\beta=0.21\pm0.013$ with the theoretical value $\beta=0.23$ for a finite size 2D XY magnet.  We  speculate that just below $T_\mathrm{N}$, the domains can exhibit any in-plane magnetic ordering direction, whereas below $T_\mathrm{EA}$ the moments fluctuate towards the in-plane easy-axes set by the Ising anisotropy. Inelastic neutron scattering does not provide evidence for interplane interactions $J_\mathrm{out}$, but only sets the upper bound of $|J_\mathrm{out}|<10^{-4}J_\mathrm{n}$ which, however, allows a lower bound on $L_\mathrm{eff}$ to be estimated as $L_\mathrm{eff}>74.5$~nm.  However, since  $L_\text{domain} < L_\mathrm{eff}$, this lower bound on $L_\mathrm{eff}$ does not provide information on the size of the domains $L_\text{domain}$. This may be the topic of a future investigation.

Finally,  we observe that BaNi$_2$V$_2$O$_8$ exhibits BKT physics. In particular, the BKT scaling accounts well for the thermal behavior of the correlation length and better than any of the other conventional models. The extracted BKT transition temperature $T_\mathrm{BKT}=44.70\pm0.25$~K falls below $T_N$, as expected for finite 2D XY systems~\cite{Bramwell, Hikami}, and its value is in overall agreement with the previously reported values of 43.3~K \cite{Heinrich} and 40.2~K \cite{Waibel2015}. The residual differences may be attributed to differences in the values of $T_N$ and the analyzed temperature regions. CMC simulations based on the Hamiltonian of BaNi$_2$V$_2$O$_8$ confirm the presence of vortex excitations, and yield $T_\mathrm{BKT}^\mathrm{CMC}=55$~K, while quantum Monte Carlo using a simplified, sign-problem free model yields $T_\mathrm{BKT}^\mathrm{QMC} = 40.2$~K respectively.

In conclusion, this comprehensive experimental and theoretical investigation identifies BaNi$_2$V$_2$O$_8$ as a rare example of an  ideal 2D magnet at all temperatures, unlike most quasi-2D magnetic compounds which instead show clear indications for 3D critical behavior. Our main achievements are (i) the development of a consistent understanding of the critical behaviour of BaNi$_2$V$_2$O$_8$ both below and above $T_N$, (ii) the identification of distinct temperature regimes where the system behaves as a finite-size 2D XY, 2D XXZ, and 2D Heisenberg antiferromagnet, (iii) the confirmation of BKT-scaling behaviour and (iv) agreement of our experimental results with classical and quantum Monte Carlo simulations using magnetic model Hamiltonians for BaNi$_2$V$_2$O$_8$.

\begin{acknowledgments}
E.K. acknowledges Ralf Feyerherm for fruitful discussions. B.L. acknowledges the support of DFG through project B06 of SFB 1143 (ID 247310070). E.K. and B.L. acknowledge ISIS neutron and muon source for the allocation of the beam time at EMU instrument. M.M. is supported by the Swedish Research Council (VR) through a Neutron Project Grant (Dnr. 2016-06955) as well as the Swedish Foundation for Strategic Research (SSF) within the Swedish national graduate school in neutron scattering (SwedNess). S.W. and L.W. acknowledge support by the DFG through Grant No. WE/3649/4-2 of the FOR 1807 and through RTG 1995. Furthermore, we thank the IT Center at RWTH Aachen University and the JSC J\"ulich for access to computing time through JARA-HPC. 
\end{acknowledgments}

\appendix
\label{appendix}

\section{Experimental details for the neutron scattering measurements}
\label{sec:Methods}
\setcounter{figure}{0} \renewcommand{\thefigure}{A.\arabic{figure}}

\begin{figure} 
\includegraphics[width=1\linewidth]{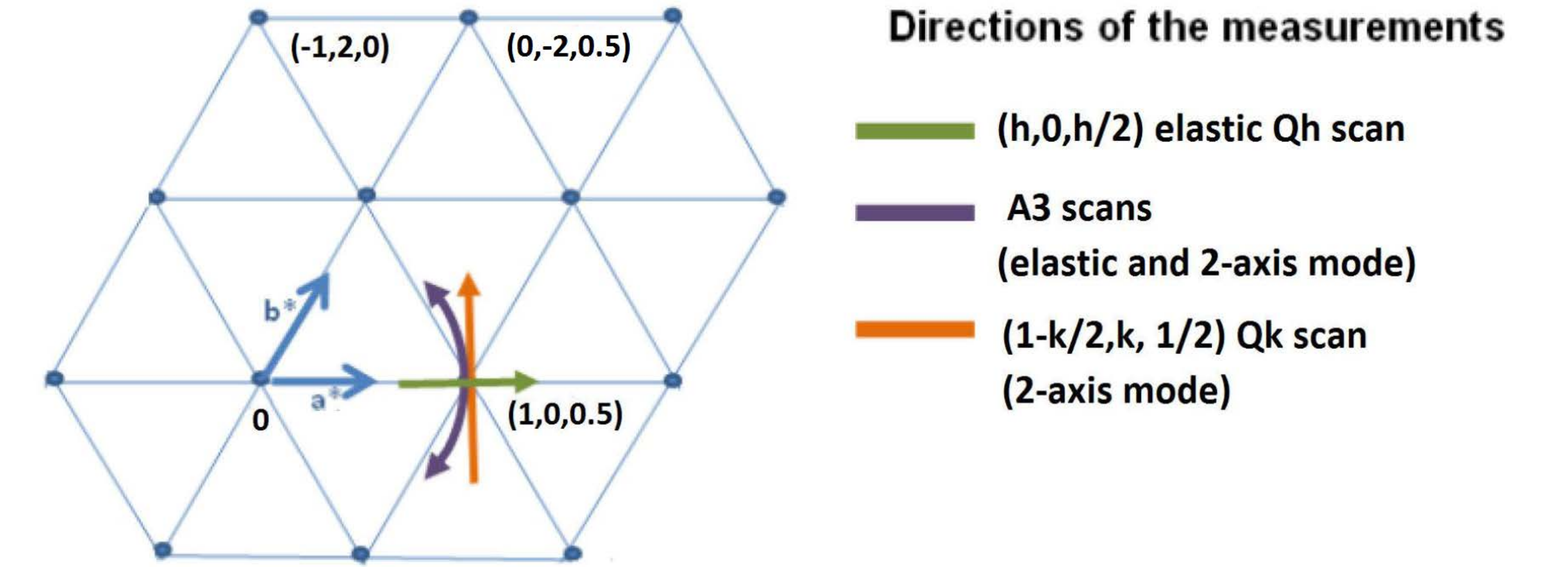} 
\caption{Sketch of the (h-$\frac{\mathrm{k}}{2}$,k,$\frac{\mathrm{h}}{2}$) scattering plane of BaNi$_2$V$_2$O$_8$ measured on TASP. The green and purple lines show the directions of the elastic $Q_\mathrm{h}$ and $A3$ scans, respectively, through the (1,0,$\frac{1}{2}$) magnetic Bragg peak. The $Q_\mathrm{k}$ and $A3$ scans for the correlation length measurement which were performed in 2-axis mode are given by the orange and purple lines respectively.} 
\label{fig:FigA1}
\end{figure}

The single crystal neutron scattering measurements of BaNi$_2$V$_2$O$_8$ were performed on the cold neutron triple-axis spectrometer, TASP, at the Paul Scherrer Institute (PSI), Switzerland. The instrument was equipped with an vertically focused Pyrolytic Graphite (002) PG(002) monochromator and a horizontally focused PG(002) analyser. A single crystal sample with a mass of 550~mg was placed inside an Orange Cryostat which cooled it down to the base temperature of $T=1.47K$~K. The measurments were performed within the (h-$\frac{\mathrm{k}}{2}$,k,$\frac{\mathrm{h}}{2}$) scattering plane which allowed the (1,0,$\frac{1}{2}$) magnetic Bragg peak to be reached. 

For the measurements of the critical exponent and ordering temperature $T_\mathrm{N}$, the analyser was set flat and the final wavevector was fixed at $k_f=1.23$~$\AA^{-1}$ providing a energy resolution of  0.074~meV which was determined by measuring the full-width-at-half-maximum (FWHM) of the elastic incoherent scattering at base temperature. To improve the statistics, elastic scans of both sample angle ($A3$) and wavevector transfer $Q_\mathrm{h}$ were performed through the (1,0,$\frac{1}{2}$) magnetic Bragg peak at many temperatures within the range 1.47-56~K.  The directions of these measurments are shown in Fig.~\ref{fig:FigA1} by the purple and green lines, respectively, where the longitudinal $Q_\mathrm{h}$ scans were perfomed along the (h,0,$\frac{\mathrm{h}}{2}$) direction. The $Q_\mathrm{h}$ and $A3$ resolution widths were found to be $\Delta Q_\mathrm{h}=0.010$~(r.l.u.) and $\Delta A3=0.468^\circ$, respectively, by fitting the FWHM of these scans 
at base temperature using the Pearson VII function. This function was found to provide the best description of the instrumental resolution function.

To measure the temperature dependence of the correlation length, the TASP spectrometer was used in two-axis diffraction mode with the analyser removed so that both elastic and inelastic signals were measured simultaneously.  A PG filter and 40' collimator were placed between the monochromator and sample and the incident wave vector was fixed at $k_i=2.662 \AA^{-1}$. $A3$-scans through the (1,0,$\frac{1}{2}$) position were measured at 1.47~K and over the temperature range from 48 to 68~K in steps of 0.25 and 1~K (purple line in Fig.~\ref{fig:FigA1}). The $A3$ angle resolution was $\Delta A3=0.387^{\circ}$ as determined from the FWHM of the scan through the (1,0,$\frac{1}{2}$) magnetic Bragg peak at base temperature.  

The correlation length was also investigated by measuring transverse $Q_\mathrm{k}$-scans through the (1,0,$\frac{1}{2}$) position (orange line in Fig.~\ref{fig:FigA1}) to improve the statistics and check the reproducibility of the results. These measurements were performed over the temperature range 48 to 140~K with steps of 0.25, 0.5, 1, 2, 5, 10 and 40~K depending on the temperature region. The TASP instrument settings were kept the same as for the $A3$-scans. The measurements were performed in the (h-$\frac{\mathrm{k}}{2}$,k,$\frac{\mathrm{h}}{2}$) scattering plane along the (1-$\frac{\mathrm{k}}{2}$,k,$\frac{1}{2}$) direction. The $Q_\mathrm{k}$ resolution was found to be $\Delta Q_\mathrm{k}=0.0102$~(r.l.u.) as determined from the FWHM of the scan at base temperature. 

To extrect the correlation length from the $Q$ and $A3$-scans collected in two-axis mode, these scans were fitted by a Lorentzian function convolved with the respective resolution function. The correlation lengths $\xi_{A3}$ and $\xi_{Q}$ were taken to be the inverse of the FWHM of this fitted Lorentzian converted to the units of inverse \AA ngstrom. $\xi_{A3}$ and $\xi_{Q}$ were found to be in good agreement with each other and were fitted simultaneously during the analysis.

\section{$T_\mathrm{N}$ from the neutron measurements}
\label{sec:TNneutrons}
\setcounter{figure}{0} \renewcommand{\thefigure}{B.\arabic{figure}}

\begin{figure} 
\includegraphics[width=1\linewidth]{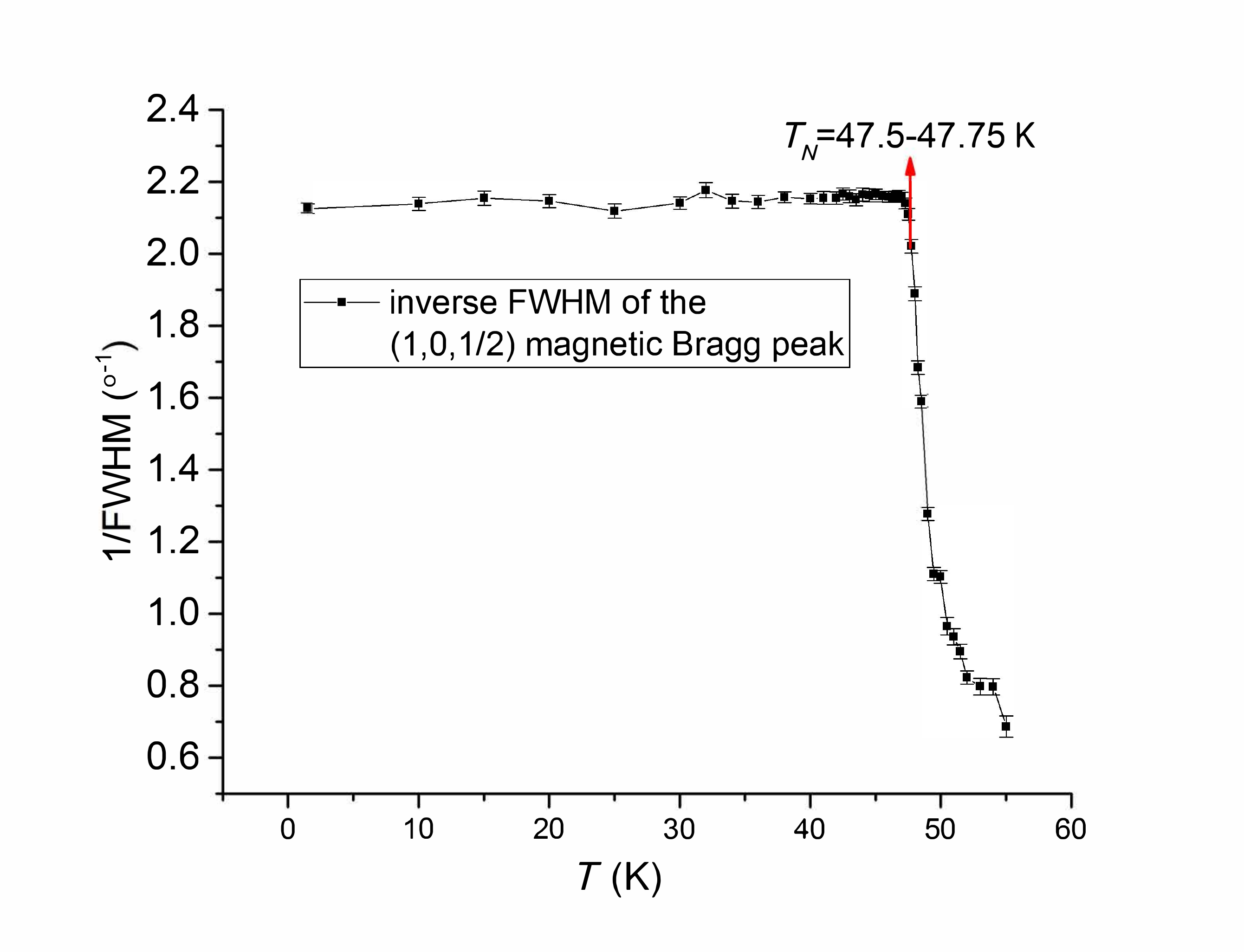} 
\caption{The inverse FWHM width of the elastic neutron $A3$-scans through the (1,0,$\frac{1}{2}$) magnetic Bragg peak plotted as a function of temperature.}
\label{fig:FigB} 
\end{figure}

When a magnetic system has long range magnetic order, its magnetic Bragg peaks are delta functions whose experimental FWHM is determined only by the resolution function. On heating, the loss of the long-range magnetic order at $T_\mathrm{N}$ leads to a finite broadening of this peak. Thus, the FWHM of the magnetic Bragg peak is a sensitive parameter to investigate the ordering temperature. 

Figure~\ref{fig:FigB} shows the inverse FWHM of the (1,0,$\frac{1}{2}$) magnetic Bragg peak of BaNi$_2$V$_2$O$_8$ plotted as a function of temperature over the range 1.5 to 56~K. The FWHM was determined by fitting the PearsonVII function to this peak at each temperature. The results reveal that the inverse FWHM is constant at finite temperatures below $T < 47.5$~K within the fitting error, while above $T=47.75$~K it sharply decreases. This suggests  that $T_\mathrm{N}= 47.75$~K. 

\section{$T_\mathrm{N}$ from $\mu^+$SR measurements in weak transverse field}
\label{sec:TNmuons}
\setcounter{figure}{0} \renewcommand{\thefigure}{C.\arabic{figure}}

\begin{figure}
\includegraphics[width=0.75\linewidth]{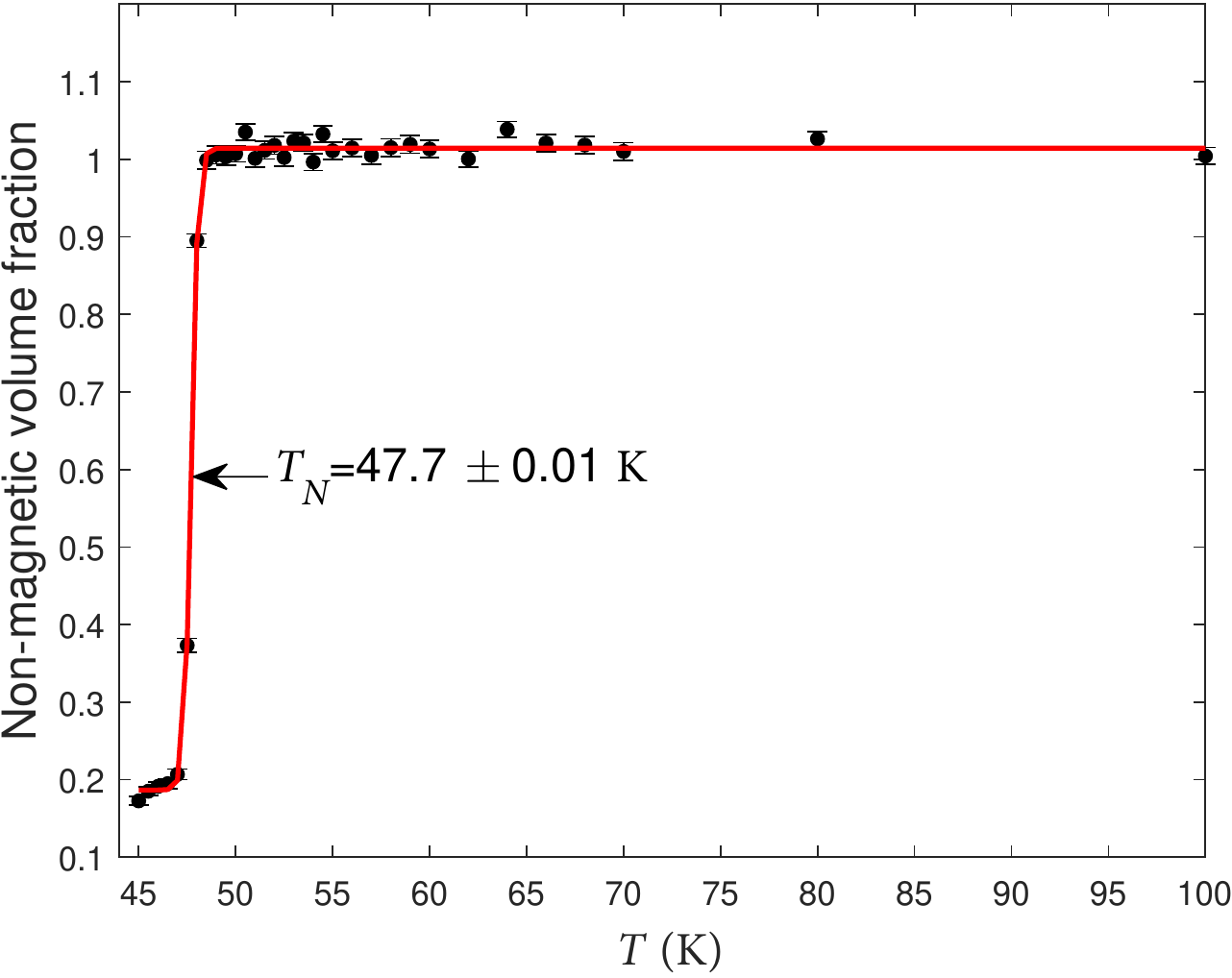}
\caption{Temperature dependence of the nonmagnetic volume fraction of the muon signal measured in  weak transverse field. The solid red line gives the best fit using Eq.~\eqref{EqTF}}
\label{fig:FigTF} 
\end{figure}

Muon spin rotation measurements were also used to determine the value of the N\'{e}el temperature. Weak transverse field (TF) $\mu^+$SR measurements were performed on a single crystal of BaNi$_2$V$_2$O$_8$ using the EMU spectrometer at the ISIS Neutron and Muon Source, UK \cite{MuonData}. The sample was oriented so that the muon beam was perpendicular to the honeycomb plane of the crystal. The data were collected over the temperature range 45-100~K in a transverse magnetic field of $B_\mathrm{TF}=20$~G. The high temperature spectra measured above $T=47.5$~K were fitted by the function \cite{TFfunction}:
\begin{equation}
A(t)=A_{\mathrm{TF}}\cdot e^{-\lambda t}\cos(\omega_{\mathrm{TF}} t+\phi)+A_{\lambda_{\mathrm{bg}}}e^{-\lambda_{\mathrm{bg}} t},
\end{equation}
where $t$ is time, $A_\mathrm{TF}$ is the amplitude of the muon spin oscillations due to the applied transverse field and $\omega_\mathrm{TF}$ is the Larmor precession frequency of these oscillations which for $B_\mathrm{TF}=20$~G is $\omega_\mathrm{TF}=0.27$~MHz. The exponential prefactor describes the damping of the oscillations with relaxation rate $\lambda$. The second non-oscillating term describes the background contribution.

At temperatures below $T_\mathrm{N}$ a second oscillation mode was clearly observed in the data which is caused by the static local internal field due to the long-range magnetic order. To account for this, the fitting function becomes
\begin{equation}
\begin{split}
A(t) &= A_{\mathrm{TF}}\cdot e^{-\lambda t}\cos(\omega_{\mathrm{TF}} t+\phi) \\
&+ A_{\mathrm{st}}\cdot e^{-\lambda_{\mathrm{st}} t}\cos(\omega_{\mathrm{st}} t+\phi_{\mathrm{st}})+A_{\lambda_{\mathrm{bg}}}e^{-\lambda_{\mathrm{bg}} t}.
\end{split}
\end{equation}
Here, $A_{\mathrm{st}}$, $\lambda_\mathrm{st}$, $\omega_{\mathrm{st}}$ and $\phi_{\mathrm{st}}$ are the muon fraction, damping, frequency and phase of the second oscillation respectively.  Figure ~\ref{fig:FigTF} shows the temperature dependence of the non-magnetic volume fraction $V(T)=\frac{A_{\mathrm{TF}(T)}}{A_{\mathrm{TF}(100)}}$ which is obtained from the extracted amplitudes $A_{\mathrm{TF}(T)}$, normalized to the amplitude $A_{\mathrm{TF}(100)}$ at the highest temperature $T=100$~K. The fraction of 18\% remaining below $T_N$ is associated with the fly-past mode used in the experiment. 
\\* To extract the transition temperature $T_{\mathrm{N}}$, the temperature dependence was fitted using the sigmoid-like function \cite{PhysRevB.82.212504,PhysRevLett.101.077001}: 
\begin{equation}
V(T)=\frac{1}{1+\exp(\frac{T_N-T}{\delta T})}+bg;
\label{EqTF}
\end{equation} 
where bg is the background and $\delta T$ describes the width of transition. The N\'{e}el  temperature was found to be $T_{\mathrm{N}}=47.7 \pm 0.01$~K which is in good agreement with the $T_{\mathrm{N}}=47.75$~K estimated from the temperature dependence of the neutron diffraction measurements.

\section{Zero-field $\mu^+$SR measurements}
\label{sec:ZFmuons}
\setcounter{figure}{0} \renewcommand{\thefigure}{D.\arabic{figure}}

\begin{figure}
\includegraphics[width=1\linewidth]{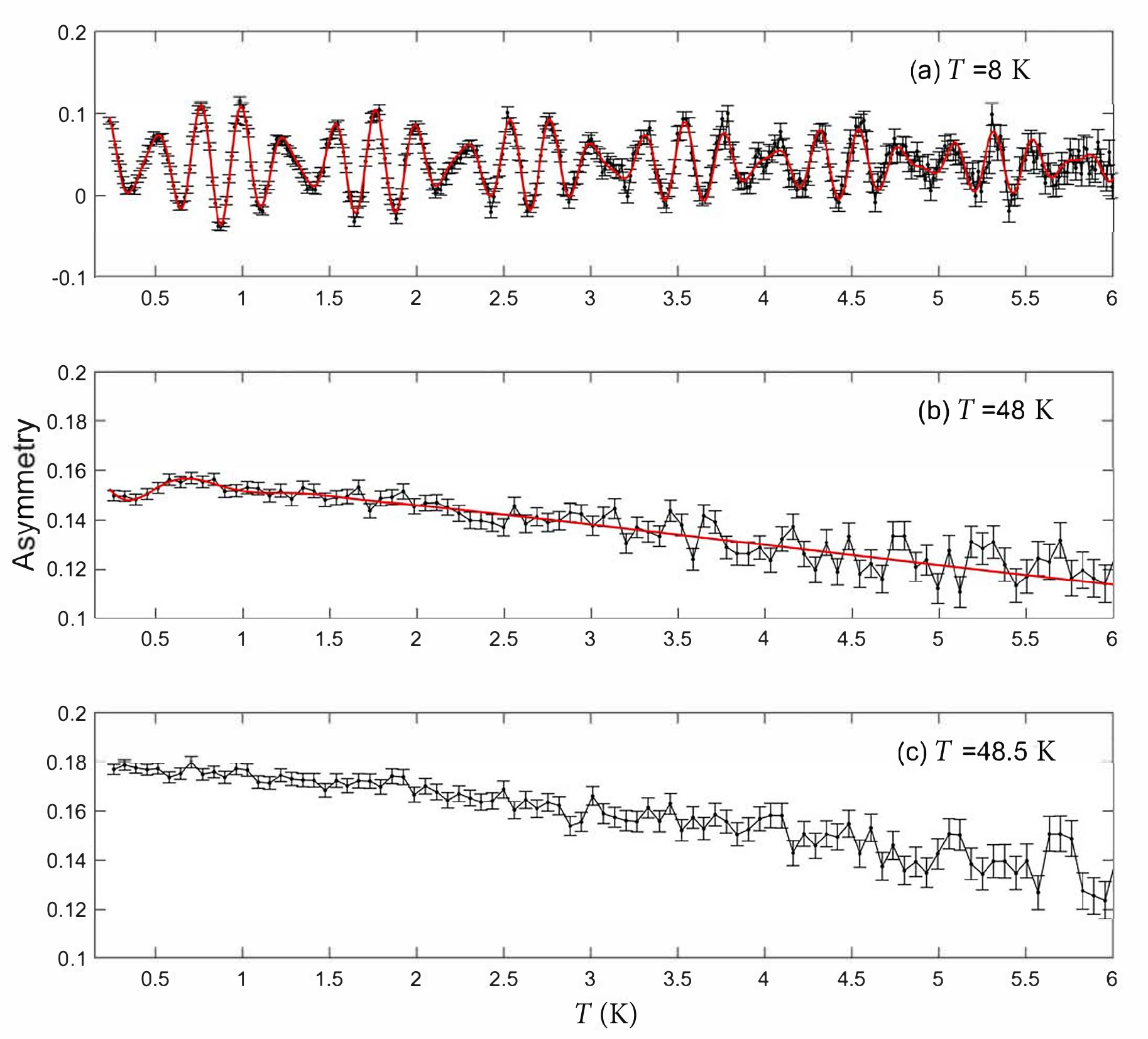} 
\caption{ZF-$\mu^+$SR spectra at (a) $T=8$~K, (b) $T=48$~K and (c) $T=48.5$~K. The single crystal sample was oriented so that the beam was parallel to the c-axis.}
\label{fig:FigM0} 
\end{figure}

Zero-field $\mu^+$SR measurements were performed on a single-crystal of BaNi$_2$V$_2$O$_8$ using the EMU spectrometer at the ISIS Neutron and Muon Source, UK \cite{MuonData}. The sample was oriented so that the muon beam was perpendicular to the honeycomb plane of the crystal and measurements took place for temperatures in the range 8-48.5~K. Figure \ref{fig:FigM0}(a) shows the ZF-$\mu^+$SR spectrum collected at $T=8$~K.  There are clear oscillations caused by the internal magnetic field of the sample due to the long-range magnetic order. The oscillations are modulated suggesting the presence of two frequencies which can be assigned to two inequivalent muon stopping sites with different internal fields. To extract these frequencies the data was fitted using the function:
\begin{equation}
{A(t) = \sum_{i=1}^{2} A_ie^{-\lambda_it}\cos(2\pi f_it)+A_{\mathrm{bg}}}
\label{eqn:muon}
\end{equation}
Here, $A_1$ and $A_2$ are the amplitudes and $f_1$ and $f_2$ are the frequencies of the two muon sites respectively. The non-oscillating term $A_{\mathrm{bg}}$ describes the background signal due to the interaction of the muons with the silver sample holder. The best fit at 8~K was achieved for $f_1=3.946\pm0.002$~MHz and $f_2=5.066\pm0.002$~MHz. These frequencies are related to the internal fields $|B_i|$, of the two muon sites via the relation $f_i=\gamma_\mu|B_i|/2\pi$ where $\gamma_\mu$ is the muon gyromagnetic ratio. 

\begin{figure} 
\includegraphics[width=0.85\linewidth]{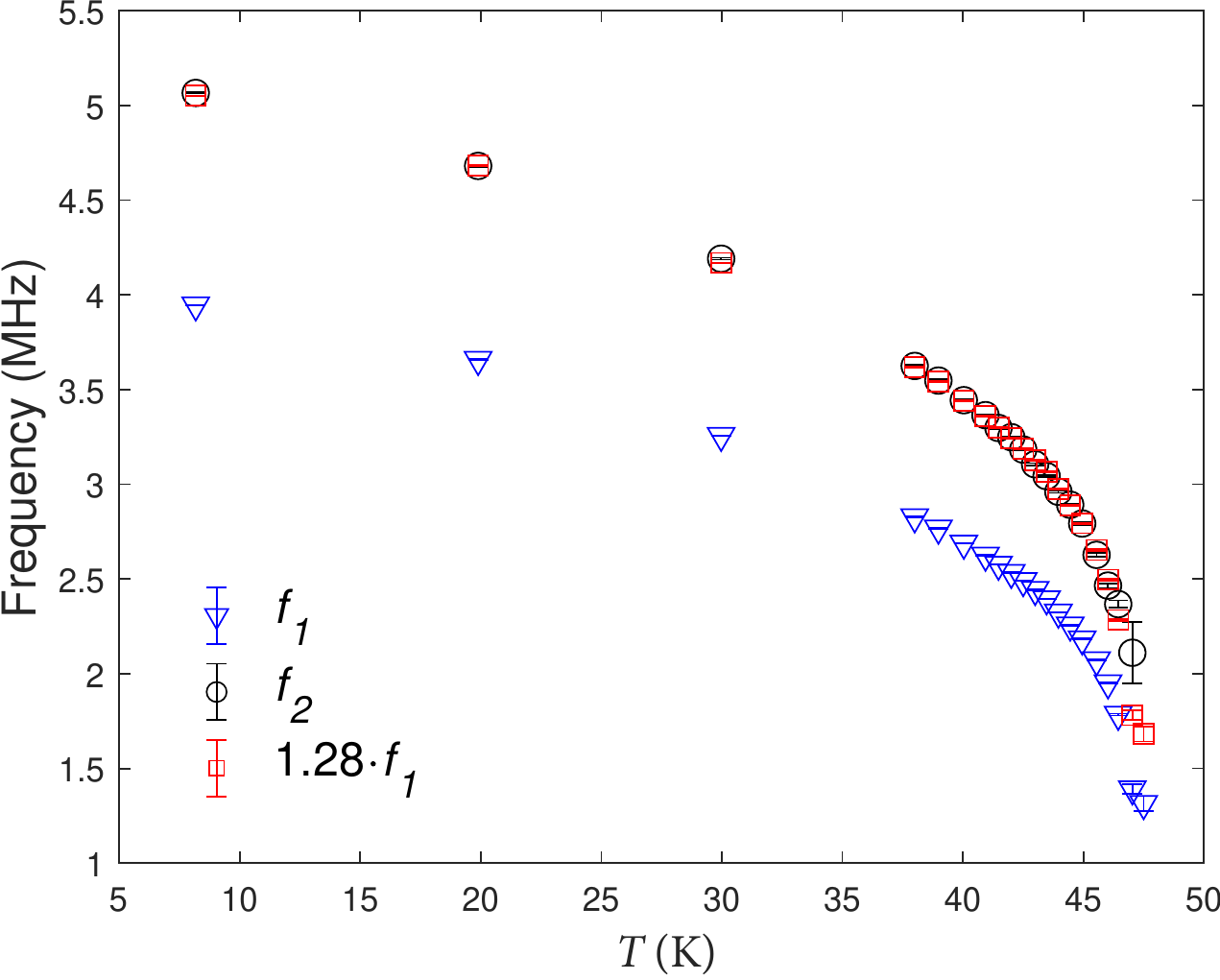} 
\caption{The temperature dependence of the two frequencies, $f_1$ (blue triangles) and $f_2$ (black circles), over the temperature range 8 - 47.5~K, extracted from the ZF-$\mu^+$SR spectra by fitting eq.\eqref{eqn:muon}. The red squares show the temperature dependence of frequency $f_1$ multiplied by the factor of 1.28.}
\label{fig:FigM} 
\end{figure} 

To explore the temperature dependence of the oscillations observed at 8~K, the  ZF-$\mu^+$SR spectra of BaNi$_2$V$_2$O$_2$ were measured at finite temperatures over the range 8 - 48.5~K. 
The extracted frequencies $f_1$ and $f_2$, are plotted as a function of temperature on Fig.~\ref{fig:FigM} where they are represented by the blue triangles and black circles respectively. Although the values of the two frequencies are different, they display the same temperature dependence up $T=46$~K suggesting that these frequencies arise from two different muon stopping sites which observe the same magnetic behavior. Indeed, as shown on Fig.~\ref{fig:FigM}, when $f_1$ is scaled by the factor 1.28, it matches $f_2$ over the temperature range 8 - 46~K.  Moreover, the amplitude ratio is found to be 2:1 which is consistent with the trigonal crystal structure of this compound.

Above $T=46$~K the frequencies display noticeably different thermal behavior and at $T=47.5$~K $f_2$ disappears. At $T=48$~K, the oscillations become almost unobservable in the spectrum (Fig.~\ref{fig:FigM0}(b)) and the fit does not converge, therefore the extracted frequency is unreliable and is excluded from Fig.\ref{fig:FigM}. At temperatures above $T=48$~K, the oscillations disappear as shown by Fig.~\ref{fig:FigM0}(c) which gives the spectrum at $T=48.5$~K.

The inconsistent thermal behavior of the frequencies above $T=46$~K can be attributed to the limitation of the muon technique in the vicinity of the transition. Indeed, the high relaxation rates of the oscillations in a critical region make the fitting of the data unreliable. Indeed, for temperatures just below $T_N$,  the internal fields are very weak and the corresponding muon oscillations have low frequencies that cannot be accurately determined due to the limited temporal resolution. Thus, for the analysis of the order parameter described in the main text, only the data below 46~K was used.

\section{Comparison of magnetization from the muon and neutron measurements}
\label{sec:Magnetization}
\setcounter{figure}{0} \renewcommand{\thefigure}{E.\arabic{figure}} 

\begin{figure}
\includegraphics[width=0.9\linewidth]{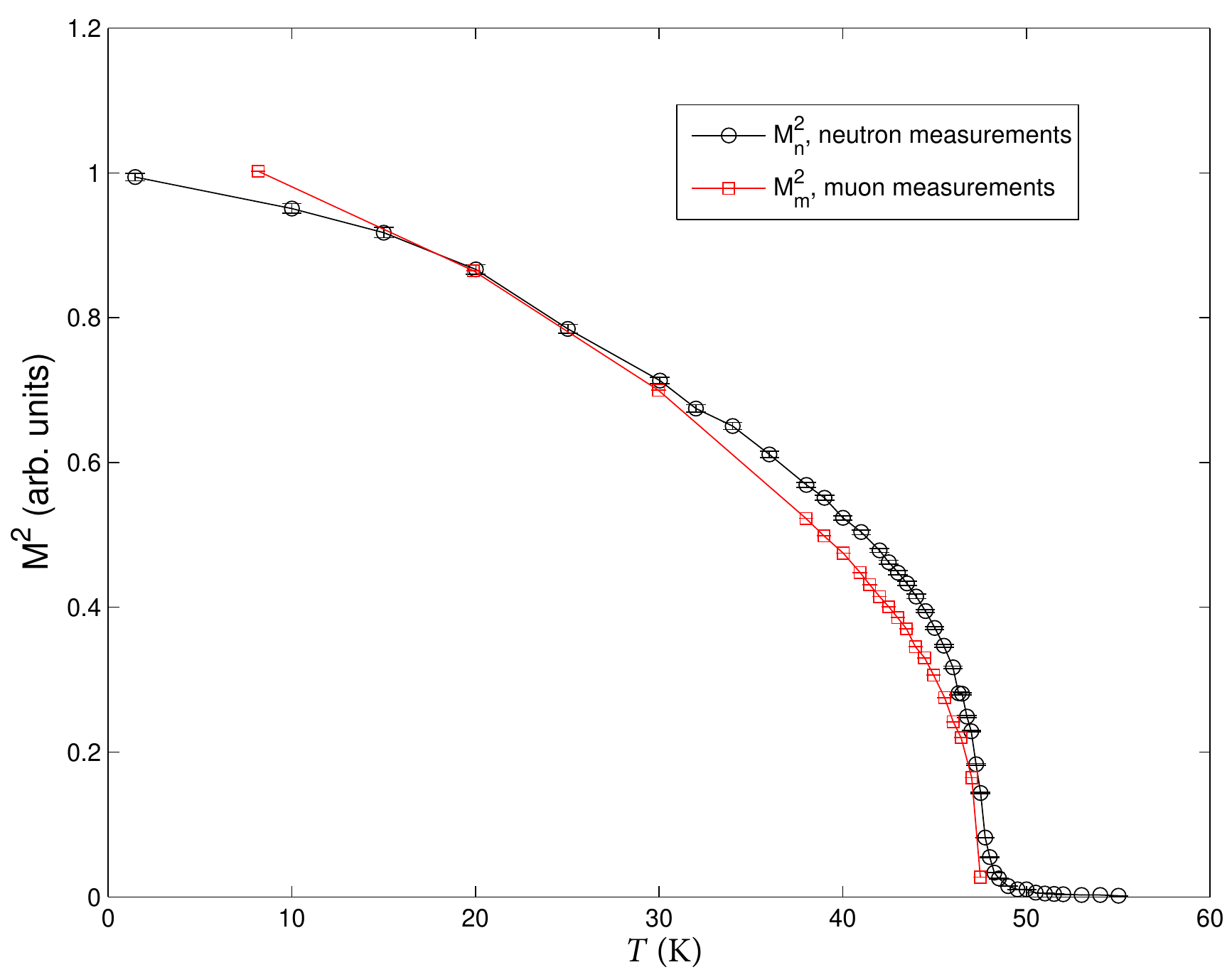} 
\caption{The temperature dependence of the square of the magnetizations $M_n^2$ and $M_m^2$, calculated from the integrated intensity of the (1,0,1/2) magnetic Bragg peak measured by neutron scattering (black circles) and from the frequencies observed in the ZF-$\mu^+$SR spectra (red squares). Lines are guides to the eye.}
\label{fig:FigA} 
\end{figure} 

The black circles on Fig.~\ref{fig:FigA} show the temperature dependence of the integrated intensity of the (1,0,1/2) magnetic Bragg peak extracted from the elastic neutron scans, which is proportional to the square of the magnetization $M^{2}_{n}$. The blue squares show the temperature dependence of the squared magnetization ${M_{m}^2}$ measured using $\mu^+$SR spectroscopy. Here, ${M_{m}^2}$ was calculated from the temperature dependence of the frequencies $f_1$ and $f_2$ observed in the ZF-$\mu^+$SR spectra. These frequencies were averaged, taking into account their respective weights. The temperature dependence of the averaged frequency $f_\mathrm{av}$ is related to the temperature dependence of the averaged internal magnetic field $|B_\mathrm{av}|$ at the muon sites. 
$|B_\mathrm{av}|$ was calculated at each temperature using the relation $2\pi f_\mathrm{av}$=$\gamma_{\mu}|B_\mathrm{av}|$, where $\gamma_\mu$ is the muon gyromagnetic ratio and ${M_{m}^2}$ was taken as ${M_{m}^2} \propto |B_\mathrm{av}|^2$. 
The values of  ${M_{m}^2}$ and ${M_n^2}$ were scaled such that they match each other at the lowest measured temperatures of 20 - 30~K. Indeed, if the system is fully static at low temperatures, then the magnetic order should be equally observed by both muon and the neutron techniques. 

The comparison of  ${M_{m}^2}$ and  ${M_n^2}$ for temperatures between 38~K and  $T_N$ reveals that the muons observe lower static fields for BaNi$_{2}$V$_{2}$O$_{8}$ than the neutrons. This difference can result from the different time scales of the muon and neutron spectroscopes. In particular, neutrons might not distinguish the slow spin-fluctuations of BaNi$_{2}$V$_{2}$O$_{8}$ and, therefore, attribute them to static signal, while muons correctly identify their dynamics.  We note that the value of ${M_{m}^2}$ is higher than that of ${M_{n}^2}$ at 8K . This indicates that the system is not fully static even at 20K -30K where the scaling was done. 
 
\section{Correlation length over the full temperature range}
\label{sec:CorrelationLength}
\setcounter{figure}{0} \renewcommand{\thefigure}{F.\arabic{figure}} 

\begin{figure}
	\includegraphics[width=0.95\linewidth]{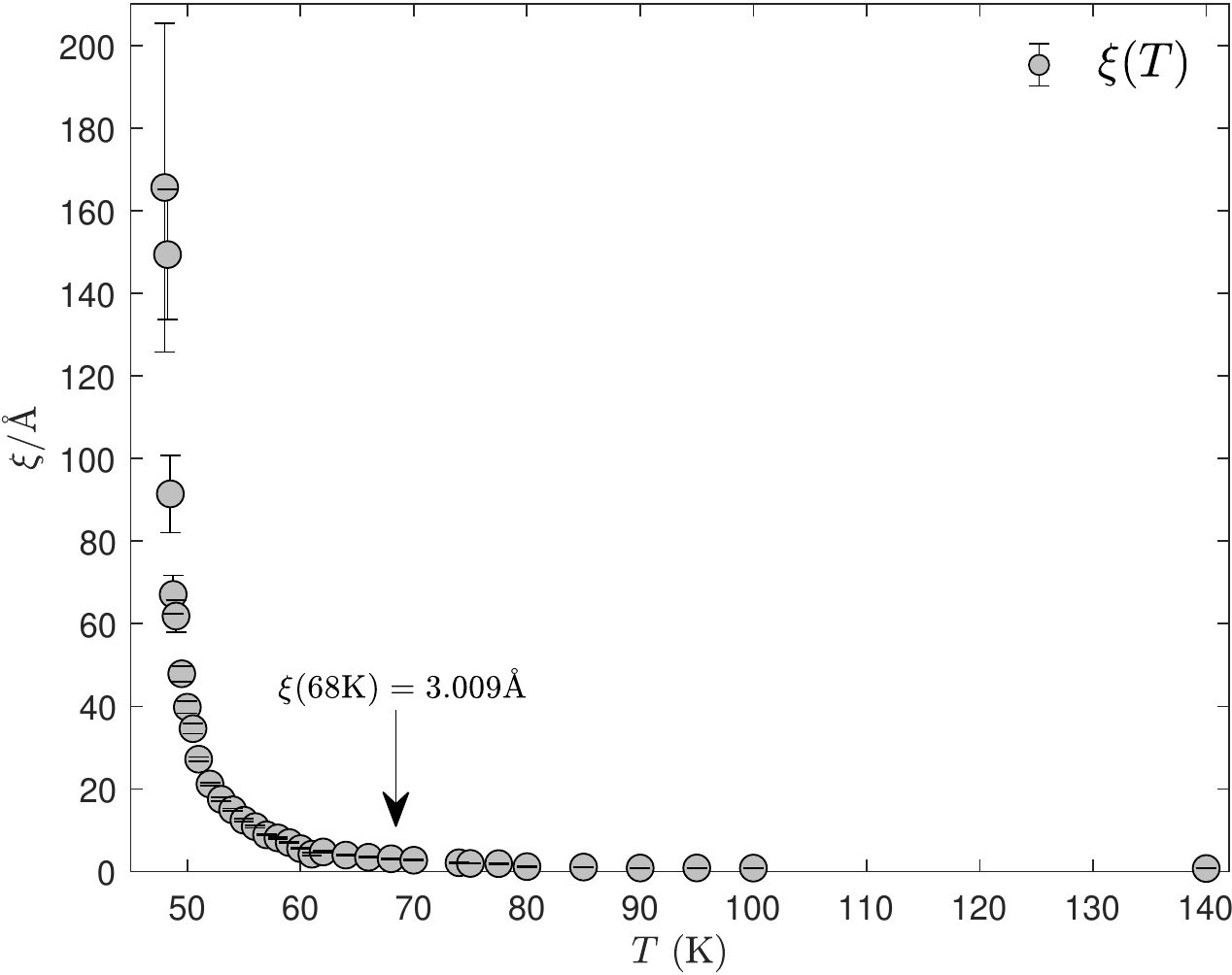}
	\caption{The correlation length $\xi(T)$, as a function of temperature $T$ up to 140~K.}
	\label{fig:FigB1} 
\end{figure}

Figure \ref{fig:FigB1} shows the correlation length, $\xi(T)$, of BaNi$_2$V$_2$O$_8$ plotted over the full temperature range up to 140~K,  as extracted from the inverse FWHM of the energy-integrated magnetic signal at wavevector ${(1,0,\frac{1}{2})}$ after taking into account the resolution broadening. At 68~K, $\xi(T)$ is comparable to the nearest neighbor in-plane Ni$^{2+}$-Ni$^{2+}$ distance, $d_{\mathrm{Ni}}=2.90$~\AA.

\section{Algebraic scaling analysis of $\xi$ on logarithmic scale}
\label{sec:AlgebraicScaling}
\setcounter{figure}{0} \renewcommand{\thefigure}{G.\arabic{figure}} 

\begin{figure}
\includegraphics[width=1\linewidth]{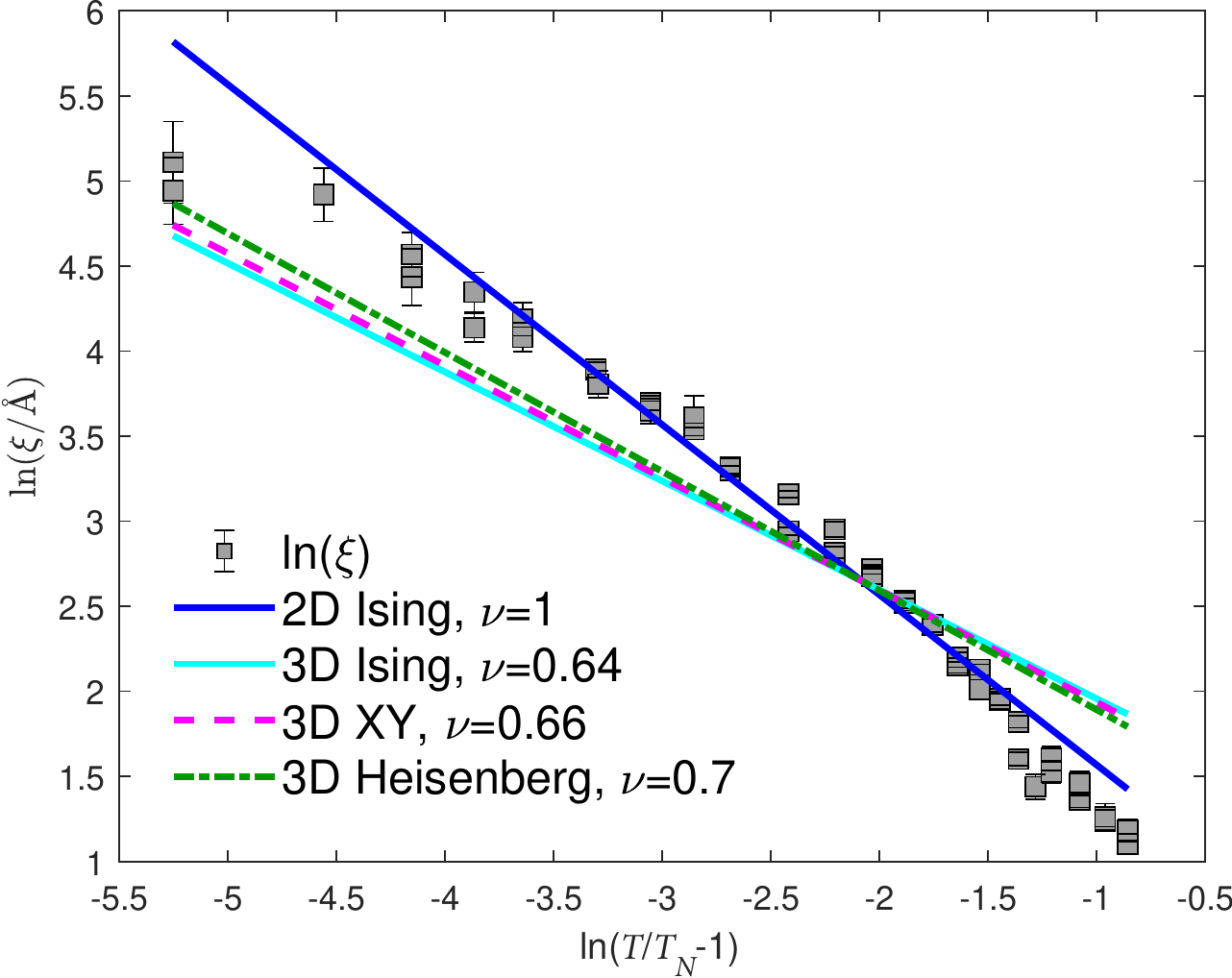}
\caption{Correlation length as a function of the reduced temperature on a logarithmic scale over the temperature 
range 48-68~K. The lines show fits to the conventional power laws 
$\nu=1$ (2D Ising, $\chi^2$=10.89), 
$\nu=0.64$ (3D Ising, $\chi^2$=57.7), 
$\nu=0.66$, (3D XY, $\chi^2$=53.43),
and $\nu=0.7$, (3D Heisenberg, $\chi^2$=45.47)
}
\label{fig:FigC1} 
\end{figure} 

In order to establish whether the correlation length of BaNi$_2$V$_2$O$_8$ follows conventional power law scaling, $\xi\propto t^{-\nu}$, the correlation length was plotted on a logarithmic scale as a function of the logarithm of the reduced temperature $t$, over the temperature range 48 - 68~K. Figure \ref{fig:FigC1} reveals that $\ln\xi$ does not follow a straight line as a function of $\ln t$, therefore no single power law scaling can describe $\xi(T)$ well.  None of the fits to the conventional power laws (2D Ising, 3D Ising, 3D XY and 3D Heisenberg) agree with the data over the entire temperature range, although the 2D Ising model gives better agreement than the others.

\section{BKT scaling compared to other 2D models.}
\label{sec:BKTScaling1}
\setcounter{figure}{0} \renewcommand{\thefigure}{H.\arabic{figure}} 

\begin{figure}
\includegraphics[width=0.9\linewidth]{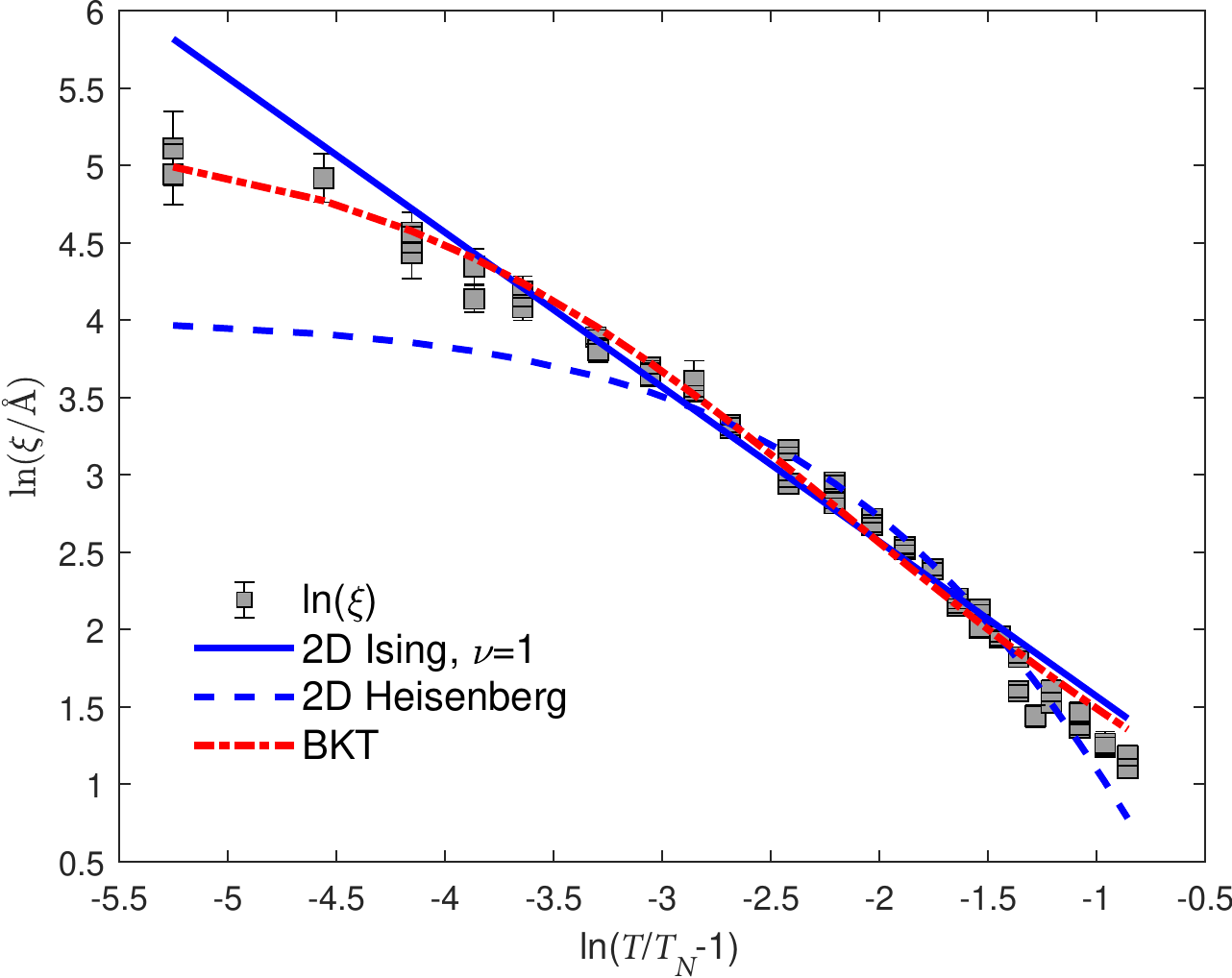} 
\caption{
Correlation length  as a function of the reduced temperature on a logarithmic scale over the temperature range 48 -68~K, and fitted to the 2D Ising model  scaling ($\nu=1$, $\chi^2$=10.89),  the 2D Heisenberg model scaling ($\chi^2$=11.93) and the BKT scaling ($\chi^2$=6.98).
}
\label{fig:FigC2} 
\end{figure}

Figure \ref{fig:FigC2} shows  $\ln\xi$ plotted as a function of $\ln t$ over the temperature range 48 - 68~K. The fit to the 2D Ising model scaling, which was found to yield a better agreement than the other conventional powers (see Appendix~\ref{sec:AlgebraicScaling} is shown. The resulting straight line noticeably deviates from the experimental data, especially close to $T_N$, for $\ln t\lesssim-4$. The fit to the 2D Heisenberg model (Eq.~\eqref{eq:1} in the main text) is also shown
It deviates strongly from the experimental data for $\ln t<-2.68$, and thus 2D Heisenberg model scaling describes $\xi$ well only for temperatures above 51~K. Finally, we include the BKT scaling formula (Eq.~\eqref{eq:2} from the main text). 
We find  that the BKT model reproduces the data over the entire explored temperature range, and especially in the vicinity of T$_N$, where neither the power laws nor the 2D Heisenberg model follow the data. 

\section{BKT scaling of $\xi(T)$ over different temperature ranges}
\label{sec:BKTScaling2}
\setcounter{figure}{0} \renewcommand{\thefigure}{I.\arabic{figure}} 

\begin{figure}
\includegraphics[width=1\linewidth]{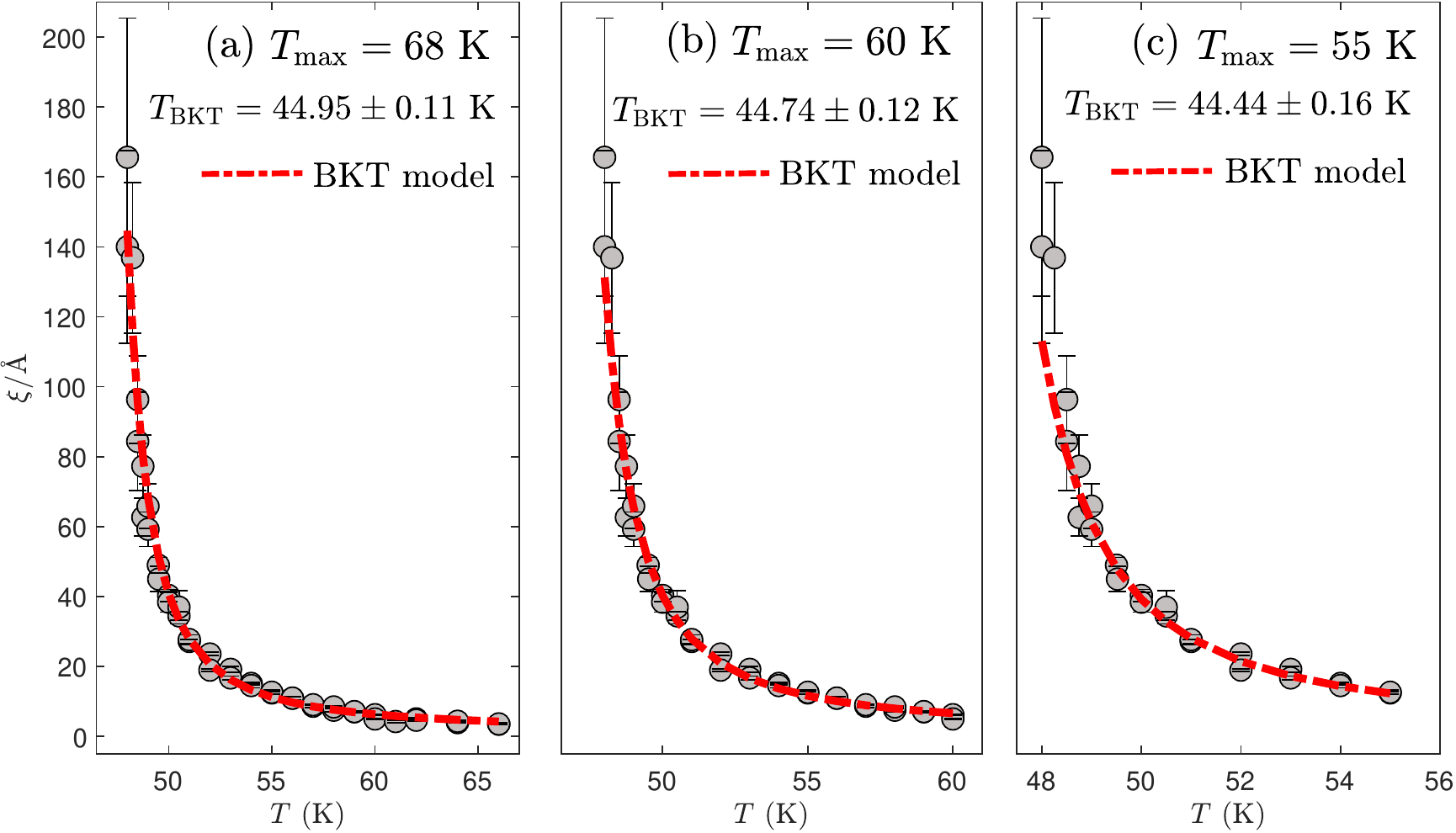} 
\caption{Correlation length of BaNi$_2$V$_2$O$_8$ fitted using the BKT formula over the temperature range from 48~K to $T_\mathrm{max}$ where (a) $T_\mathrm{max}=66$~K (b) $T_\mathrm{max}=60$~K (c) $T_\mathrm{max}=55$~K.}
\label{fig:FigE}  
\end{figure}

The correlation length were analyzed using BKT theory over several temperature regions, from 48~K to $T_\mathrm{max}$, where $T_\mathrm{max}=66$, 60 and 55~K,  to assess the sensitivity of  the extracted value of $T_\mathrm{BKT}$ on  the  temperature range used in the fitting. The results are presented on Fig.\ref{fig:FigE} and reveal that  T$_{BKT}$ lies within the range of 44.44~K$<T_\mathrm{BKT}<44.95$~K. Therefore, $T_\mathrm{BKT}$ is fairly insensitive to the explored temperature range and can be averaged to the value $T_\mathrm{BKT}=44.70 \pm 0.25$~K for further analysis.

\section{Details of Classical Monte Carlo simulations}
\label{sec:CMC}
\setcounter{figure}{0} \renewcommand{\thefigure}{J.\arabic{figure}} 

For our classical Monte Carlo calculations we use a standard single-spin update Metropolis algorithm for a lattice with $N=1560$ honeycomb sites and periodic boundary conditions. After a sufficiently long equilibration time the real space spin configurations, spin correlations  and magnetic susceptibility $\chi^\mu$ ($\mu=x,y,z$) are obtained from the numerical outputs for different temperatures $T$. The susceptibility $\chi^\mu$ is calculated from
\begin{equation}
\chi^\mu(T)=\frac{1}{k_\text{B}T}\langle (M^\mu-\langle M^\mu\rangle)^2\rangle\;,
\end{equation}
where $M^\mu$ is the $\mu$-component of the magnetization $M^\mu=\sum_{i=1}^N S_{\mathbf{r}_i}^\mu$. To eliminate statistical noise, the susceptibility is averaged over 400000 Monte Carlo steps (where one step consists of $N$ single-spin updates). Likewise, the spin correlations
\begin{equation}
C^\mu(\mathbf{r})=\frac{1}{N}\sum_{i=1}^N\langle S_{\mathbf{r}_i}^\mu S_{\mathbf{r}_i+\mathbf{r}}^\mu \rangle
\end{equation}
are calculated as a function of the distance $r=|\mathbf{r}|$ and the resulting correlation function $C^\mu(r)$ is fitted against an exponential decay $\sim e^{-r/\xi}$ for large temperatures (above the BKT temperature) and an algebraic decay $\sim r^{-\eta}$ for small temperatures (below the BKT temperature). The temperature $T=55K$ where the inplane correlation functions $C^x(r)$, $C^y(r)$ show an exponent $\eta=1/4$ and the correlations change from an algebraic to an exponential behavior is identified as the BKT transition temperature. At selected Monte Carlo times and for various different temperatures, snapshots of the real-space spin configurations are analyzed with respect to the occurrence of vortices, see Fig. 4(a)-(c) of the main text. For each hexagon of the honeycomb lattice, we consider the azimuthal angles (i.e., inplane spin orientations) $\phi_a$ ($a=0,1,\ldots,5$) for the six adjacent honeycomb sites. To find the winding number $w$ of a possible vortex located at this hexagon we calculate the differences $\Delta \phi_a=\phi_{a+1}-\phi_{a}$ (with $\phi_6\equiv\phi_0$) which, due to the $2\pi$-periodic property of azimuthal angles, can be defined such that they obey $-\pi<\Delta \phi_a\leq \pi$. The winding number $w$ associated with the spin configuration around a hexagon is then given by $w=\sum_{a=0}^5\Delta\phi_a/(2\pi)$. In Fig. 4(a)-(c) of the main text, we mark a vortex with $w=1$ ($w=-1$) by a closed (open) sphere.

\section{Details of Quantum Monte Carlo simulations}
\label{sec:QMC}
\setcounter{figure}{0} \renewcommand{\thefigure}{K.\arabic{figure}} 

For the QMC simulations, we used the stochastic series expansion quantum Monte Carlo method with the directed loop update~\cite{PhysRevB.59.R14157,PhysRevB.43.5950,PhysRevB.62.1102} for the Hamiltonian 
\begin{equation}
H = J^\mathrm{QMC}_n \sum_{\langle i,j\rangle} S_i \cdot S_j + \sum_i h^\text{ani}_i.
\end{equation}
The parallel susceptibility $\chi_{\parallel c}$ was measured by introducing the anisotropy
\begin{equation}
h_i^\text{ani} = D^\mathrm{QMC}_\mathrm{EP(XY)} (S_i^z)^2,
\end{equation}
which is diagonal in the standard $S_z$ computational basis.
In order to access $\chi_{\perp c}$, the introduced anisotropy was
\begin{eqnarray}
h_i^\text{ani} &=& D^\mathrm{QMC}_\mathrm{EP(XY)} (S_i^x)^2\\
& =& \frac{D^\mathrm{QMC}_\mathrm{EP(XY)}}{4} \left((S_i^+)^2 + S^+_i S^-_i + S^-_i S^+_i + (S^-_i)^2\right),\nonumber
\end{eqnarray}
which is off-diagonal, but can still be sampled without a sign problem within the framework of the directed loop update. This global spin rotation allows us to readily measure both susceptibilities in the $S_z$ basis. We find that the reported results are converged to the thermodynamic limit within the statistical error bars for $L=42$.

\section{Scaling of the QMC simulations to the experimental data}
\label{sec:QMCvsData}
\setcounter{figure}{0} \renewcommand{\thefigure}{L.\arabic{figure}} 

The QMC computations provide the magnetic susceptibility $\chi^\mathrm{QMC}_\mathrm{red} (\zeta)$ in terms of the dimensionless parameter $\zeta=k_B T / J^\mathrm{QMC}_n$, which is scaled to 
compare to the experimental data to
\begin{equation}
 \chi^\mathrm{QMC}(T)=\frac{N_A \:  (g^\mathrm{QMC})^2\:  \mu^2_B}{J^\mathrm{QMC}_n}\chi^\mathrm{QMC}_\mathrm{red} (k_B T/J^\mathrm{QMC}_n)+\chi_\mathrm{dia},
 \end{equation}
where $N_A$ is the Avogadro number, $g$ is the $g$-factor, $\mu_B$ the Bohr magneton and $\chi_\mathrm{dia}$ a constant associated with the diamagnetic contribution. The best agreement with the experimental data was achieved for $J^\mathrm{QMC}_n=8.07$ meV, $g^\mathrm{QMC}_{||c}=2.07$, $g^\mathrm{QMC}_{\perp c}=2.17$ and $\chi_\mathrm{dia}$ of  order  $10^{-4}$~cm$^{3}$/mol~Ni. These $g$-factors are similar to  the experimentally measured values of $g_{||c}=2.225$ and $g_{\perp c}=2.243$ \cite{Heinrich}.

\section{Results of QMC computations for the Hamiltonian without the D$_\mathrm{EP(XY)}$ term}
\label{sec:QMCEasyPlane}
\setcounter{figure}{0} \renewcommand{\thefigure}{M.\arabic{figure}} 

\begin{figure}
	\includegraphics[width=0.9\linewidth]{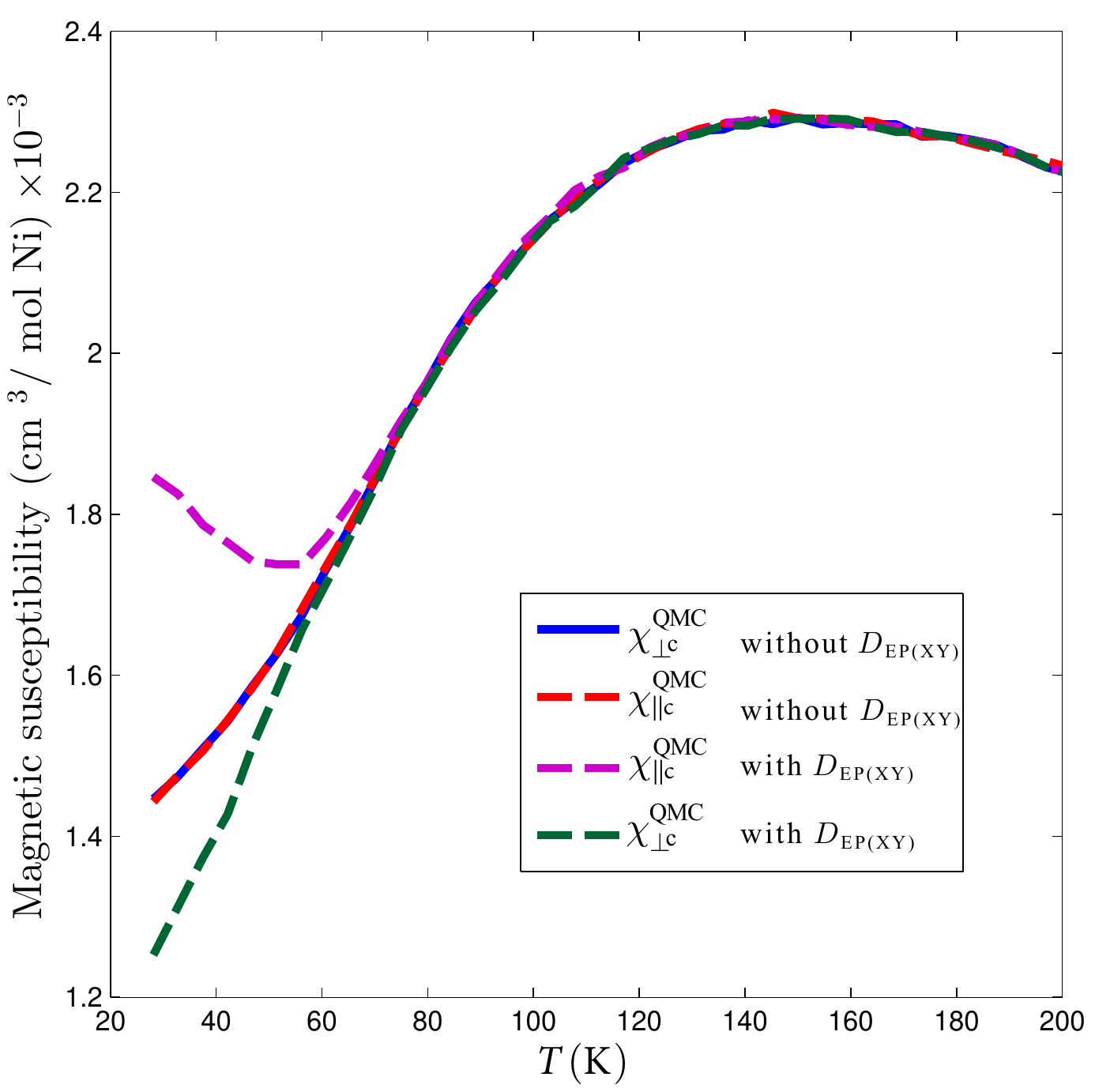} 
	\caption{Magnetic susceptibility of BaNi$_2$V$_2$O$_8$ parallel and perpendicular to the c-axis obtained by QMC simulations with  and without the  
		$D_\mathrm{EP(XY)}$ term.}
	\label{fig:FigD} 
\end{figure}

Fig.~\ref{fig:FigD} presents the QMC simulations of the magnetic susceptibility parallel and perpendicular to the c-axis computed for the Hamiltonian of BaNi$_2$V$_2$O$_8$ with and without the D$_\mathrm{EP(XY)}$ term. The results reveal isotropic behavior for the magnetic susceptibility computed without the anisotropy term over the entire temperature range. 

In contrast, the magnetic susceptibility computed for the Hamiltonian with planar anisotropy reveals strongly anisotropic behavior. In particular, the magnetic susceptibility computed parallel to the $c$-axis has a characteristic minimum. Thus, these computations confirm that the term $D_\mathrm{EP(XY)}$ is responsible for the anisotropy and, also, for the minimum at $T_\mathrm{XY} = 51$ K observed in the $\chi^\mathrm{QMC}_{||c}$. Therefore, the minimum at $T_\mathrm{XY}$ observed in the experimental susceptibility can be associated with the crossover to the XY-dominated  regime.

\section{Simplified Hamiltonian of BaNi$_2$V$_2$O$_8$} 
\label{sec:SimpHam}
\setcounter{figure}{0} \renewcommand{\thefigure}{N.\arabic{figure}} 

\begin{figure*}
	\includegraphics[width=1\linewidth]{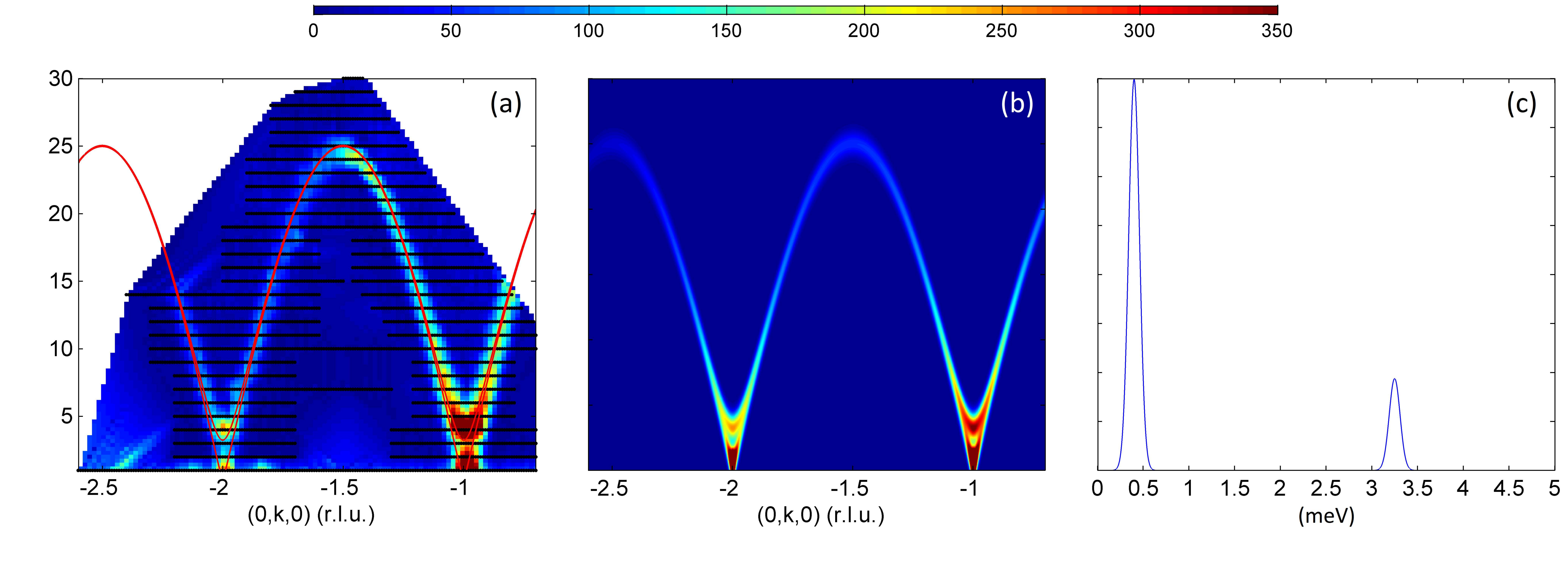} 
	\caption{Single crystal magnetic excitation spectrum of BaNi$_2$V$_2$O$_8$ along the (0,k,0) direction (a) measured at $T=3.5$~K \cite{KlyushinaBaNi} and (b) computed using the spin-wave theory. (c) Computed energy scan at $Q$=(1,0,0). The calculations used the Hamiltonian (Eq.~\eqref{eq:Ham}( with parameters $J_n=8.8$~meV, $J_\mathrm{out}= -0.00045$~meV, $D_\mathrm{EP(XY)}=0.099$~meV and $D_\mathrm{EA}=-0.0014$~meV.}
	\label{fig:FigC} 
\end{figure*}

The magnetic excitation spectrum of BaNi$_2$V$_2$O$_8$ was measured at low temperatures in the magnetically ordered phase using inelastic neutron scattering. The data were used to obtain the Hamiltonian by fitting it to spin-wave theory, where  the first three intraplane nearest neighbour interactions, the interplane interaction, the easy-plane anisotropy and the weak in-plane easy-axis single-ion anisotropy of the Ni$^{2+}$ magnetic ions were considered. The instrument settings of these experiments as well as the data analysis are discussed in Ref.~\cite{KlyushinaBaNi}. 

Because a simplified Hamiltonian was used for the QMC calculations, the spectrum was refitted to verify the accuracy of this Hamiltonian. Fig.~\ref{fig:FigC}(a) shows the measured spin-waves along the (0,k,0) direction while Fig.~\ref{fig:FigC}(b) shows the corresponding spectrum computed using the SpinW MatLab library \cite{SpinW} for the simplified Hamiltonian:
\begin{equation} 
\begin{split}
H =\displaystyle  J_{n} \sum_{\langle i,j\rangle}\boldsymbol{S}_i \cdot \boldsymbol{S}_j+  \displaystyle J_\mathrm{out}\sum_{\langle i,j\rangle'} \boldsymbol{S}_i \cdot \boldsymbol{S}_j \\ 
+\displaystyle\sum_{i} D_\mathrm{EP(XY)} (S_i^{c})^2+ \displaystyle\sum_{i} D_\mathrm{EA} (S_i^{x})^2
\label{eq:Ham}
\end{split}
\end{equation}
Here, $J_n$ and $J_\mathrm{out}$ are the first-neighbor intraplane and interplane magnetic exchange couplings, while $D_\mathrm{EP(XY)}$ and $D_\mathrm{EA}$ are the easy-plane and in-plane easy-axis single-ion anisotropies, respectively. The simulations were performed for all three twins and the results were averaged. The values $J_n=8.8$~meV, $J_\mathrm{out}=-0.00045$~meV, $D_\mathrm{EP(XY)}=0.099$~meV and $D_\mathrm{EA}=-0.0014$~meV provided the best agreement with the data (see Fig.~\ref{fig:FigC}(b)). In particular, $J_n$ is responsible for the energy scale of the dispersion shown in Fig.~\ref{fig:FigC}(a), while the parameters $D_\mathrm{EP(XY)}$ and $D_\mathrm{EA}$ generate the energy gaps at the antiferromagnetic zone center. The sizes of these gaps were extracted by fitting the experimental data corrected for resolution effects and were found to be $E_1=0.41$~meV and $E_2=3.25$~meV \cite{KlyushinaBaNi}. Figure~\ref{fig:FigC}(c) presents the energy scan at $Q$=(1,0,0) computed for the best fit parameters which reproduces both the gaps. The extracted values $J_n=8.8$~meV and $D_\mathrm{EP(XY)}=0.099$~meV are very similar to the ones obtained by fitting the QMC simulations to the susceptibility data. 
%


\begin{thebibliography}{42}%
\makeatletter
\providecommand \@ifxundefined [1]{%
 \@ifx{#1\undefined}
}%
\providecommand \@ifnum [1]{%
 \ifnum #1\expandafter \@firstoftwo
 \else \expandafter \@secondoftwo
 \fi
}%
\providecommand \@ifx [1]{%
 \ifx #1\expandafter \@firstoftwo
 \else \expandafter \@secondoftwo
 \fi
}%
\providecommand \natexlab [1]{#1}%
\providecommand \enquote  [1]{``#1''}%
\providecommand \bibnamefont  [1]{#1}%
\providecommand \bibfnamefont [1]{#1}%
\providecommand \citenamefont [1]{#1}%
\providecommand \href@noop [0]{\@secondoftwo}%
\providecommand \href [0]{\begingroup \@sanitize@url \@href}%
\providecommand \@href[1]{\@@startlink{#1}\@@href}%
\providecommand \@@href[1]{\endgroup#1\@@endlink}%
\providecommand \@sanitize@url [0]{\catcode `\\12\catcode `\$12\catcode
  `\&12\catcode `\#12\catcode `\^12\catcode `\_12\catcode `\%12\relax}%
\providecommand \@@startlink[1]{}%
\providecommand \@@endlink[0]{}%
\providecommand \url  [0]{\begingroup\@sanitize@url \@url }%
\providecommand \@url [1]{\endgroup\@href {#1}{\urlprefix }}%
\providecommand \urlprefix  [0]{URL }%
\providecommand \Eprint [0]{\href }%
\providecommand \doibase [0]{http://dx.doi.org/}%
\providecommand \selectlanguage [0]{\@gobble}%
\providecommand \bibinfo  [0]{\@secondoftwo}%
\providecommand \bibfield  [0]{\@secondoftwo}%
\providecommand \translation [1]{[#1]}%
\providecommand \BibitemOpen [0]{}%
\providecommand \bibitemStop [0]{}%
\providecommand \bibitemNoStop [0]{.\EOS\space}%
\providecommand \EOS [0]{\spacefactor3000\relax}%
\providecommand \BibitemShut  [1]{\csname bibitem#1\endcsname}%
\let\auto@bib@innerbib\@empty
\bibitem [{\citenamefont {Kosterlitz}\ and\ \citenamefont
  {Thouless}(1973)}]{KosterlitzThouless}%
  \BibitemOpen
  \bibfield  {author} {\bibinfo {author} {\bibfnamefont {J~M}\ \bibnamefont
  {Kosterlitz}}\ and\ \bibinfo {author} {\bibfnamefont {D~J}\ \bibnamefont
  {Thouless}},\ }\bibfield  {title} {\enquote {\bibinfo {title} {Ordering,
  metastability and phase transitions in two-dimensional systems},}\ }\href
  {\doibase 10.1088/0022-3719/6/7/010} {\bibfield  {journal} {\bibinfo
  {journal} {Journal of Physics C: Solid State Physics}\ }\textbf {\bibinfo
  {volume} {6}},\ \bibinfo {pages} {1181--1203} (\bibinfo {year}
  {1973})}\BibitemShut {NoStop}%
\bibitem [{\citenamefont {Kosterlitz}(1974)}]{Kosterlitz}%
  \BibitemOpen
  \bibfield  {author} {\bibinfo {author} {\bibfnamefont {J~M}\ \bibnamefont
  {Kosterlitz}},\ }\bibfield  {title} {\enquote {\bibinfo {title} {The critical
  properties of the two-dimensional xy model},}\ }\href {\doibase
  10.1088/0022-3719/7/6/005} {\bibfield  {journal} {\bibinfo  {journal}
  {Journal of Physics C: Solid State Physics}\ }\textbf {\bibinfo {volume}
  {7}},\ \bibinfo {pages} {1046--1060} (\bibinfo {year} {1974})}\BibitemShut
  {NoStop}%
\bibitem [{\citenamefont {Mermin}\ and\ \citenamefont {Wagner}(1966)}]{Mermin}%
  \BibitemOpen
  \bibfield  {author} {\bibinfo {author} {\bibfnamefont {N.~D.}\ \bibnamefont
  {Mermin}}\ and\ \bibinfo {author} {\bibfnamefont {H.}~\bibnamefont
  {Wagner}},\ }\bibfield  {title} {\enquote {\bibinfo {title} {Absence of
  ferromagnetism or antiferromagnetism in one- or two-dimensional isotropic
  heisenberg models},}\ }\href {\doibase 10.1103/PhysRevLett.17.1133}
  {\bibfield  {journal} {\bibinfo  {journal} {Phys. Rev. Lett.}\ }\textbf
  {\bibinfo {volume} {17}},\ \bibinfo {pages} {1133--1136} (\bibinfo {year}
  {1966})}\BibitemShut {NoStop}%
\bibitem [{\citenamefont {Berezinskii}(1971)}]{BerezinskiiP1}%
  \BibitemOpen
  \bibfield  {author} {\bibinfo {author} {\bibfnamefont {V.~L.}\ \bibnamefont
  {Berezinskii}},\ }\bibfield  {title} {\enquote {\bibinfo {title} {Destruction
  of long-range order in one-dimensional and two-dimensional systems having a
  continuous symmetry group i. classical systems},}\ }\href
  {http://jetp.ac.ru/cgi-bin/dn/e_032_03_0493.pdf} {\bibfield  {journal}
  {\bibinfo  {journal} {JETP}\ }\textbf {\bibinfo {volume} {32}},\ \bibinfo
  {pages} {493} (\bibinfo {year} {1971})}\BibitemShut {NoStop}%
\bibitem [{\citenamefont {Berezinskii}(1972)}]{BerezinskiiP2}%
  \BibitemOpen
  \bibfield  {author} {\bibinfo {author} {\bibfnamefont {V.~L.}\ \bibnamefont
  {Berezinskii}},\ }\bibfield  {title} {\enquote {\bibinfo {title} {Destruction
  of long-range order in one-dimensional and two-dimensional systems possessing
  a continuous symmetry group. ii. quantum systems},}\ }\href
  {http://www.jetp.ac.ru/cgi-bin/dn/e_034_03_0610.pdf} {\bibfield  {journal}
  {\bibinfo  {journal} {JETP}\ }\textbf {\bibinfo {volume} {34}},\ \bibinfo
  {pages} {610} (\bibinfo {year} {1972})}\BibitemShut {NoStop}%
\bibitem [{\citenamefont {Xu}\ \emph {et~al.}(2000)\citenamefont {Xu},
  \citenamefont {Broholm}, \citenamefont {Reich},\ and\ \citenamefont
  {Adams}}]{Xu}%
  \BibitemOpen
  \bibfield  {author} {\bibinfo {author} {\bibfnamefont {Guangyong}\
  \bibnamefont {Xu}}, \bibinfo {author} {\bibfnamefont {C.}~\bibnamefont
  {Broholm}}, \bibinfo {author} {\bibfnamefont {Daniel~H.}\ \bibnamefont
  {Reich}}, \ and\ \bibinfo {author} {\bibfnamefont {M.~A.}\ \bibnamefont
  {Adams}},\ }\bibfield  {title} {\enquote {\bibinfo {title} {Triplet waves in
  a quantum spin liquid},}\ }\href {\doibase 10.1103/PhysRevLett.84.4465}
  {\bibfield  {journal} {\bibinfo  {journal} {Phys. Rev. Lett.}\ }\textbf
  {\bibinfo {volume} {84}},\ \bibinfo {pages} {4465--4468} (\bibinfo {year}
  {2000})}\BibitemShut {NoStop}%
\bibitem [{\citenamefont {Schneider}\ \emph {et~al.}(2014)\citenamefont
  {Schneider}, \citenamefont {Zaitsev}, \citenamefont {Fuchs},\ and\
  \citenamefont {von L{\"o}hneysen}}]{BKTthinFilms}%
  \BibitemOpen
  \bibfield  {author} {\bibinfo {author} {\bibfnamefont {R.}~\bibnamefont
  {Schneider}}, \bibinfo {author} {\bibfnamefont {A.~G.}\ \bibnamefont
  {Zaitsev}}, \bibinfo {author} {\bibfnamefont {D.}~\bibnamefont {Fuchs}}, \
  and\ \bibinfo {author} {\bibfnamefont {H}~\bibnamefont {von L{\"o}hneysen}},\
  }\bibfield  {title} {\enquote {\bibinfo {title} {Excess conductivity and
  {B}erezinskii{\textendash}{K}osterlitz{\textendash}{T}houless transition in
  superconducting {FeSe} thin films},}\ }\href {\doibase
  10.1088/0953-8984/26/45/455701} {\bibfield  {journal} {\bibinfo  {journal}
  {Journal of Physics: Condensed Matter}\ }\textbf {\bibinfo {volume} {26}},\
  \bibinfo {pages} {455701} (\bibinfo {year} {2014})}\BibitemShut {NoStop}%
\bibitem [{\citenamefont {Tutsch}\ \emph {et~al.}(2014)\citenamefont {Tutsch},
  \citenamefont {Wolf}, \citenamefont {Wessel}, \citenamefont {Postulka},
  \citenamefont {Tsui}, \citenamefont {Jeschke}, \citenamefont {Opahle},
  \citenamefont {Saha-Dasgupta}, \citenamefont {Valenti}, \citenamefont
  {Br{\"u}hl}, \citenamefont {Removic-Langer}, \citenamefont {Kretz},
  \citenamefont {Lerner}, \citenamefont {Wagner},\ and\ \citenamefont
  {Lang}}]{BKTfield}%
  \BibitemOpen
  \bibfield  {author} {\bibinfo {author} {\bibfnamefont {U.}~\bibnamefont
  {Tutsch}}, \bibinfo {author} {\bibfnamefont {B.}~\bibnamefont {Wolf}},
  \bibinfo {author} {\bibfnamefont {S.}~\bibnamefont {Wessel}}, \bibinfo
  {author} {\bibfnamefont {L.}~\bibnamefont {Postulka}}, \bibinfo {author}
  {\bibfnamefont {Y.}~\bibnamefont {Tsui}}, \bibinfo {author} {\bibfnamefont
  {H.O.}\ \bibnamefont {Jeschke}}, \bibinfo {author} {\bibfnamefont
  {I.}~\bibnamefont {Opahle}}, \bibinfo {author} {\bibfnamefont
  {T.}~\bibnamefont {Saha-Dasgupta}}, \bibinfo {author} {\bibfnamefont
  {R.}~\bibnamefont {Valenti}}, \bibinfo {author} {\bibfnamefont
  {A.}~\bibnamefont {Br{\"u}hl}}, \bibinfo {author} {\bibfnamefont
  {K.}~\bibnamefont {Removic-Langer}}, \bibinfo {author} {\bibfnamefont
  {T.}~\bibnamefont {Kretz}}, \bibinfo {author} {\bibfnamefont {H.-W.}\
  \bibnamefont {Lerner}}, \bibinfo {author} {\bibfnamefont {M.}~\bibnamefont
  {Wagner}}, \ and\ \bibinfo {author} {\bibfnamefont {M.}~\bibnamefont
  {Lang}},\ }\bibfield  {title} {\enquote {\bibinfo {title} {Evidence of a
  field-induced {B}erezinskii-{K}osterlitz-{T}houless scenario in a
  two-dimensional spin–dimer system},}\ }\href {\doibase 10.1038/ncomms6169}
  {\bibfield  {journal} {\bibinfo  {journal} {Nature Comm.}\ }\textbf {\bibinfo
  {volume} {5}},\ \bibinfo {pages} {6169} (\bibinfo {year} {2014})}\BibitemShut
  {NoStop}%
\bibitem [{\citenamefont {Opherden}\ \emph {et~al.}(2020)\citenamefont
  {Opherden}, \citenamefont {Nizar}, \citenamefont {Richardson}, \citenamefont
  {Monroe}, \citenamefont {Turnbull}, \citenamefont {Polson}, \citenamefont
  {Vela}, \citenamefont {Blackmore}, \citenamefont {Goddard}, \citenamefont
  {Singleton}, \citenamefont {Choi}, \citenamefont {Xiao}, \citenamefont
  {Williams}, \citenamefont {Lancaster}, \citenamefont {Pratt}, \citenamefont
  {Blundell}, \citenamefont {Skourski}, \citenamefont {Uhlarz}, \citenamefont
  {Ponomaryov}, \citenamefont {Zvyagin}, \citenamefont {Wosnitza},
  \citenamefont {Baenitz}, \citenamefont {Heinmaa}, \citenamefont {Stern},
  \citenamefont {K\"uhne},\ and\ \citenamefont
  {Landee}}]{opherden2020extremely}%
  \BibitemOpen
  \bibfield  {author} {\bibinfo {author} {\bibfnamefont {D.}~\bibnamefont
  {Opherden}}, \bibinfo {author} {\bibfnamefont {N.}~\bibnamefont {Nizar}},
  \bibinfo {author} {\bibfnamefont {K.}~\bibnamefont {Richardson}}, \bibinfo
  {author} {\bibfnamefont {J.~C.}\ \bibnamefont {Monroe}}, \bibinfo {author}
  {\bibfnamefont {M.~M.}\ \bibnamefont {Turnbull}}, \bibinfo {author}
  {\bibfnamefont {M.}~\bibnamefont {Polson}}, \bibinfo {author} {\bibfnamefont
  {S.}~\bibnamefont {Vela}}, \bibinfo {author} {\bibfnamefont {W.~J.~A.}\
  \bibnamefont {Blackmore}}, \bibinfo {author} {\bibfnamefont {P.~A.}\
  \bibnamefont {Goddard}}, \bibinfo {author} {\bibfnamefont {J.}~\bibnamefont
  {Singleton}}, \bibinfo {author} {\bibfnamefont {E.~S.}\ \bibnamefont {Choi}},
  \bibinfo {author} {\bibfnamefont {F.}~\bibnamefont {Xiao}}, \bibinfo {author}
  {\bibfnamefont {R.~C.}\ \bibnamefont {Williams}}, \bibinfo {author}
  {\bibfnamefont {T.}~\bibnamefont {Lancaster}}, \bibinfo {author}
  {\bibfnamefont {F.~L.}\ \bibnamefont {Pratt}}, \bibinfo {author}
  {\bibfnamefont {S.~J.}\ \bibnamefont {Blundell}}, \bibinfo {author}
  {\bibfnamefont {Y.}~\bibnamefont {Skourski}}, \bibinfo {author}
  {\bibfnamefont {M.}~\bibnamefont {Uhlarz}}, \bibinfo {author} {\bibfnamefont
  {A.~N.}\ \bibnamefont {Ponomaryov}}, \bibinfo {author} {\bibfnamefont
  {S.~A.}\ \bibnamefont {Zvyagin}}, \bibinfo {author} {\bibfnamefont
  {J.}~\bibnamefont {Wosnitza}}, \bibinfo {author} {\bibfnamefont
  {M.}~\bibnamefont {Baenitz}}, \bibinfo {author} {\bibfnamefont
  {I.}~\bibnamefont {Heinmaa}}, \bibinfo {author} {\bibfnamefont
  {R.}~\bibnamefont {Stern}}, \bibinfo {author} {\bibfnamefont
  {H.}~\bibnamefont {K\"uhne}}, \ and\ \bibinfo {author} {\bibfnamefont
  {C.~P.}\ \bibnamefont {Landee}},\ }\bibfield  {title} {\enquote {\bibinfo
  {title} {Extremely well isolated two-dimensional spin-$\frac{1}{2}$
  antiferromagnetic {H}eisenberg layers with a small exchange coupling in the
  molecular-based magnet {C}u{POF}},}\ }\href {\doibase
  10.1103/PhysRevB.102.064431} {\bibfield  {journal} {\bibinfo  {journal}
  {Phys. Rev. B}\ }\textbf {\bibinfo {volume} {102}},\ \bibinfo {pages}
  {064431} (\bibinfo {year} {2020})}\BibitemShut {NoStop}%
	\bibitem [{\citenamefont {Hu}\ \emph {et~al.}(2020)\citenamefont {Hu},
  \citenamefont {Ma}, \citenamefont {Liao}, \citenamefont {Li}, \citenamefont
  {Ma}, \citenamefont {Cui}, \citenamefont {Shangguan}, \citenamefont {Huang},
  \citenamefont {Qi}, \citenamefont {Li}, \citenamefont {Meng}, \citenamefont
  {Wen},\ and\ \citenamefont {Yu}}]{BKTIsing}%
  \BibitemOpen
  \bibfield  {author} {\bibinfo {author} {\bibfnamefont {Z.}~\bibnamefont
  {Hu}}, \bibinfo {author} {\bibfnamefont {Z.}~\bibnamefont {Ma}}, \bibinfo
  {author} {\bibfnamefont {Y.-D.}\ \bibnamefont {Liao}}, \bibinfo {author}
  {\bibfnamefont {H.}~\bibnamefont {Li}}, \bibinfo {author} {\bibfnamefont
  {C.}~\bibnamefont {Ma}}, \bibinfo {author} {\bibfnamefont {Y.}~\bibnamefont
  {Cui}}, \bibinfo {author} {\bibfnamefont {Y.}~\bibnamefont {Shangguan}},
  \bibinfo {author} {\bibfnamefont {Y.}~\bibnamefont {Huang}}, \bibinfo
  {author} {\bibfnamefont {Y.}~\bibnamefont {Qi}}, \bibinfo {author}
  {\bibfnamefont {W.}~\bibnamefont {Li}}, \bibinfo {author} {\bibfnamefont
  {Z.~Y.}\ \bibnamefont {Meng}}, \bibinfo {author} {\bibfnamefont
  {J.}~\bibnamefont {Wen}}, \ and\ \bibinfo {author} {\bibfnamefont
  {W.}~\bibnamefont {Yu}},\ }\bibfield  {title} {\enquote {\bibinfo {title}
  {Evidence of the berezinskii-kosterlitz-thouless phase in a frustrated
  magnet},}\ }\href {\doibase 10.1038/s41467-020-19380-x} {\bibfield  {journal}
  {\bibinfo  {journal} {Nature Communications}\ }\textbf {\bibinfo {volume}
  {11}},\ \bibinfo {pages} {5631} (\bibinfo {year} {2020})}\BibitemShut
  {NoStop}%
\bibitem [{\citenamefont {Hirakawa}(1982)}]{K2CuF4Hirakawa}%
  \BibitemOpen
  \bibfield  {author} {\bibinfo {author} {\bibfnamefont {K.}~\bibnamefont
  {Hirakawa}},\ }\bibfield  {title} {\enquote {\bibinfo {title}
  {{K}osterlitz-{T}houless transition in two-dimensional planar ferromagnet
  {K}$_2${C}u{F}$_4$ (invited)},}\ }\href {\doibase 10.1063/1.330706}
  {\bibfield  {journal} {\bibinfo  {journal} {J. Appl. Phys}\ }\textbf
  {\bibinfo {volume} {53}},\ \bibinfo {pages} {1893--1898} (\bibinfo {year}
  {1982})}\BibitemShut {NoStop}%
\bibitem [{\citenamefont {Als-Nielsen}\ \emph {et~al.}(1993)\citenamefont
  {Als-Nielsen}, \citenamefont {Bramwell}, \citenamefont {Hutchings},
  \citenamefont {McIntyre},\ and\ \citenamefont {Visser}}]{Als_Nielsen_1993}%
  \BibitemOpen
  \bibfield  {author} {\bibinfo {author} {\bibfnamefont {J.}~\bibnamefont
  {Als-Nielsen}}, \bibinfo {author} {\bibfnamefont {S.~T.}\ \bibnamefont
  {Bramwell}}, \bibinfo {author} {\bibfnamefont {M.~T.}\ \bibnamefont
  {Hutchings}}, \bibinfo {author} {\bibfnamefont {G.~J.}\ \bibnamefont
  {McIntyre}}, \ and\ \bibinfo {author} {\bibfnamefont {D.}~\bibnamefont
  {Visser}},\ }\bibfield  {title} {\enquote {\bibinfo {title} {Neutron
  scattering investigation of the static critical properties of
  {R}b$_2${C}r{Cl}$_4$},}\ }\href {\doibase 10.1088/0953-8984/5/42/009}
  {\bibfield  {journal} {\bibinfo  {journal} {J. Phys. Condens. Matter}\
  }\textbf {\bibinfo {volume} {5}},\ \bibinfo {pages} {7871--7892} (\bibinfo
  {year} {1993})}\BibitemShut {NoStop}%
\bibitem [{\citenamefont {Regnault}\ and\ \citenamefont
  {Rossat-Mignod}()}]{Jongh}%
  \BibitemOpen
  \bibfield  {author} {\bibinfo {author} {\bibfnamefont {L.~P.}\ \bibnamefont
  {Regnault}}\ and\ \bibinfo {author} {\bibfnamefont {J.}~\bibnamefont
  {Rossat-Mignod}},\ }\href@noop {} {\enquote {\bibinfo {title} {Magnetic
  properties of layered transition metal compounds},}\ }\bibinfo {note}
  {(Kluwer Academic Publishers, Netherlands, 1990)}\BibitemShut {NoStop}%
\bibitem [{\citenamefont {Regnault}\ \emph {et~al.}(1983)\citenamefont
  {Regnault}, \citenamefont {Rossat-Mignod}, \citenamefont {Henry},\ and\
  \citenamefont {{de Jongh}}}]{a13}%
  \BibitemOpen
  \bibfield  {author} {\bibinfo {author} {\bibfnamefont {L.P.}\ \bibnamefont
  {Regnault}}, \bibinfo {author} {\bibfnamefont {J.}~\bibnamefont
  {Rossat-Mignod}}, \bibinfo {author} {\bibfnamefont {J.Y.}\ \bibnamefont
  {Henry}}, \ and\ \bibinfo {author} {\bibfnamefont {L.J.}\ \bibnamefont {{de
  Jongh}}},\ }\bibfield  {title} {\enquote {\bibinfo {title} {Magnetic
  properties of the quasi-2d easy plane antiferromagnet
  {B}a{N}i$_2$({PO}$_4$)$_2$},}\ }\href {\doibase
  https://doi.org/10.1016/0304-8853(83)90864-8} {\bibfield  {journal} {\bibinfo
   {journal} {Journal of Magnetism and Magnetic Materials}\ }\textbf {\bibinfo
  {volume} {31-34}},\ \bibinfo {pages} {1205 -- 1206} (\bibinfo {year}
  {1983})}\BibitemShut {NoStop}%
\bibitem [{\citenamefont {Gaveau}\ \emph {et~al.}(1991)\citenamefont {Gaveau},
  \citenamefont {Boucher}, \citenamefont {Regnault},\ and\ \citenamefont
  {Henry}}]{Gaveau}%
  \BibitemOpen
  \bibfield  {author} {\bibinfo {author} {\bibfnamefont {P.}~\bibnamefont
  {Gaveau}}, \bibinfo {author} {\bibfnamefont {J.~P.}\ \bibnamefont {Boucher}},
  \bibinfo {author} {\bibfnamefont {L.~P.}\ \bibnamefont {Regnault}}, \ and\
  \bibinfo {author} {\bibfnamefont {Y.}~\bibnamefont {Henry}},\ }\bibfield
  {title} {\enquote {\bibinfo {title} {Magnetic-field dependence of the
  phosphorus nuclear spin-relaxation rate in the quasi-two-dimensional {XY}
  antiferromagnet {B}a{N}i$_2$({PO}$_4$)$_2$},}\ }\href {\doibase
  10.1063/1.348816} {\bibfield  {journal} {\bibinfo  {journal} {J. Appl. Phys}\
  }\textbf {\bibinfo {volume} {69}},\ \bibinfo {pages} {6228--6230} (\bibinfo
  {year} {1991})}\BibitemShut {NoStop}%
\bibitem [{\citenamefont {R{\o}nnow}\ \emph {et~al.}(2000)\citenamefont
  {R{\o}nnow}, \citenamefont {Wildes},\ and\ \citenamefont
  {Bramwell}}]{RonnowMnPS3}%
  \BibitemOpen
  \bibfield  {author} {\bibinfo {author} {\bibfnamefont {H.~M.}\ \bibnamefont
  {R{\o}nnow}}, \bibinfo {author} {\bibfnamefont {A.~R.}\ \bibnamefont
  {Wildes}}, \ and\ \bibinfo {author} {\bibfnamefont {S.~T.}\ \bibnamefont
  {Bramwell}},\ }\bibfield  {title} {\enquote {\bibinfo {title} {Magnetic
  correlations in the {2D} {S}=52 honeycomb antiferromagnet {M}n{PS}$_3$},}\
  }\href {\doibase https://doi.org/10.1016/S0921-4526(99)01520-3} {\bibfield
  {journal} {\bibinfo  {journal} {Physica B: Condensed Matter}\ }\textbf
  {\bibinfo {volume} {276-278}},\ \bibinfo {pages} {676 -- 677} (\bibinfo
  {year} {2000})}\BibitemShut {NoStop}%
\bibitem [{\citenamefont {Yamada}(1972)}]{Yamada_Cv}%
  \BibitemOpen
  \bibfield  {author} {\bibinfo {author} {\bibfnamefont {Isao}\ \bibnamefont
  {Yamada}},\ }\bibfield  {title} {\enquote {\bibinfo {title} {Magnetic
  properties of {K}$_2${C}u{F}$_4$ –a transparent two-dimensional
  ferromagnet},}\ }\href {\doibase 10.1143/JPSJ.33.979} {\bibfield  {journal}
  {\bibinfo  {journal} {J. Phys. Soc. Jpn}\ }\textbf {\bibinfo {volume} {33}},\
  \bibinfo {pages} {979--988} (\bibinfo {year} {1972})}\BibitemShut {NoStop}%
\bibitem [{\citenamefont {Bloembergen}(1976)}]{BLOEMBERGEN}%
  \BibitemOpen
  \bibfield  {author} {\bibinfo {author} {\bibfnamefont {P.}~\bibnamefont
  {Bloembergen}},\ }\bibfield  {title} {\enquote {\bibinfo {title} {On the
  specific heat of some layered copper compounds: {II}. {M}agnetic
  contribution},}\ }\href {\doibase
  https://doi.org/10.1016/0378-4363(76)90098-X} {\bibfield  {journal} {\bibinfo
   {journal} {Physica B+C}\ }\textbf {\bibinfo {volume} {85}},\ \bibinfo
  {pages} {51 -- 72} (\bibinfo {year} {1976})}\BibitemShut {NoStop}%
\bibitem [{\citenamefont {Regnault}\ \emph {et~al.}(1980)\citenamefont
  {Regnault}, \citenamefont {Henry}, \citenamefont {Rossat-Mignod},\ and\
  \citenamefont {De~Combarieu}}]{Regnault3}%
  \BibitemOpen
  \bibfield  {author} {\bibinfo {author} {\bibfnamefont {L.~P.}\ \bibnamefont
  {Regnault}}, \bibinfo {author} {\bibfnamefont {J.~Y.}\ \bibnamefont {Henry}},
  \bibinfo {author} {\bibfnamefont {J.}~\bibnamefont {Rossat-Mignod}}, \ and\
  \bibinfo {author} {\bibfnamefont {A.}~\bibnamefont {De~Combarieu}},\
  }\bibfield  {title} {\enquote {\bibinfo {title} {Magnetic properties of the
  layered nickel compounds {B}a{N}i$_2$({PO}$_4$)$_2$ and
  {B}a{N}i$_2$({A}s{O}$_4$)$_2$},}\ }\href {\doibase
  https://doi.org/10.1016/0304-8853(80)90869-0} {\bibfield  {journal} {\bibinfo
   {journal} {Journal of Magnetism and Magnetic Materials}\ }\textbf {\bibinfo
  {volume} {15-18}},\ \bibinfo {pages} {1021 -- 1022} (\bibinfo {year}
  {1980})}\BibitemShut {NoStop}%
\bibitem [{\citenamefont {Hirakawa}\ and\ \citenamefont
  {Ikeda}(1973)}]{K2CuF4crit}%
  \BibitemOpen
  \bibfield  {author} {\bibinfo {author} {\bibfnamefont {K.}~\bibnamefont
  {Hirakawa}}\ and\ \bibinfo {author} {\bibfnamefont {H.}~\bibnamefont
  {Ikeda}},\ }\bibfield  {title} {\enquote {\bibinfo {title} {Investigations of
  two-dimensional ferromagnet {K}$_2${C}u{F}$_4$ by neutron scattering},}\
  }\href {\doibase 10.1143/JPSJ.35.1328} {\bibfield  {journal} {\bibinfo
  {journal} {J. Phys. Soc. Jpn}\ }\textbf {\bibinfo {volume} {35}},\ \bibinfo
  {pages} {1328--1336} (\bibinfo {year} {1973})}\BibitemShut {NoStop}%
\bibitem [{\citenamefont {Kleemann}\ and\ \citenamefont
  {Sch{\"a}fer}(1983)}]{KLEEMANN}%
  \BibitemOpen
  \bibfield  {author} {\bibinfo {author} {\bibfnamefont {W.}~\bibnamefont
  {Kleemann}}\ and\ \bibinfo {author} {\bibfnamefont {F.J.}\ \bibnamefont
  {Sch{\"a}fer}},\ }\bibfield  {title} {\enquote {\bibinfo {title} {Critical
  behavior of the magnetization of {K}$_2${C}u{F}$_4$ and {R}b$_2${C}r{C}l$_4$
  a comparative magneto-optical study},}\ }\href {\doibase
  https://doi.org/10.1016/0304-8853(83)90579-6} {\bibfield  {journal} {\bibinfo
   {journal} {J. Magn. Magn. Mat}\ }\textbf {\bibinfo {volume} {31-34}},\
  \bibinfo {pages} {565 -- 566} (\bibinfo {year} {1983})}\BibitemShut {NoStop}%
\bibitem [{\citenamefont {Wildes}\ \emph {et~al.}(2006)\citenamefont {Wildes},
  \citenamefont {R{\o}nnow}, \citenamefont {Roessli}, \citenamefont {Harris},\
  and\ \citenamefont {Godfrey}}]{Wildes2006}%
  \BibitemOpen
  \bibfield  {author} {\bibinfo {author} {\bibfnamefont {A.~R.}\ \bibnamefont
  {Wildes}}, \bibinfo {author} {\bibfnamefont {H.~M.}\ \bibnamefont
  {R{\o}nnow}}, \bibinfo {author} {\bibfnamefont {B.}~\bibnamefont {Roessli}},
  \bibinfo {author} {\bibfnamefont {M.~J.}\ \bibnamefont {Harris}}, \ and\
  \bibinfo {author} {\bibfnamefont {K.~W.}\ \bibnamefont {Godfrey}},\
  }\bibfield  {title} {\enquote {\bibinfo {title} {Static and dynamic critical
  properties of the quasi-two-dimensional antiferromagnet
  ${\mathrm{mnps}}_{3}$},}\ }\href {\doibase 10.1103/PhysRevB.74.094422}
  {\bibfield  {journal} {\bibinfo  {journal} {Phys. Rev. B}\ }\textbf {\bibinfo
  {volume} {74}},\ \bibinfo {pages} {094422} (\bibinfo {year}
  {2006})}\BibitemShut {NoStop}%
\bibitem [{\citenamefont {Hikami}\ and\ \citenamefont
  {Tsuneto}(1980)}]{Hikami}%
  \BibitemOpen
  \bibfield  {author} {\bibinfo {author} {\bibfnamefont {S.}~\bibnamefont
  {Hikami}}\ and\ \bibinfo {author} {\bibfnamefont {T.}~\bibnamefont
  {Tsuneto}},\ }\bibfield  {title} {\enquote {\bibinfo {title} {Phase
  transition of quasi-two dimensional planar system},}\ }\href {\doibase
  10.1143/PTP.63.387} {\bibfield  {journal} {\bibinfo  {journal} {Prog. Theor.
  Phys.}\ }\textbf {\bibinfo {volume} {63}},\ \bibinfo {pages} {387} (\bibinfo
  {year} {1980})}\BibitemShut {NoStop}%
\bibitem [{\citenamefont {Bramwell}\ and\ \citenamefont
  {Holdsworth}(1993)}]{Bramwell}%
  \BibitemOpen
  \bibfield  {author} {\bibinfo {author} {\bibfnamefont {S.~T.}\ \bibnamefont
  {Bramwell}}\ and\ \bibinfo {author} {\bibfnamefont {P.~C.~W.}\ \bibnamefont
  {Holdsworth}},\ }\bibfield  {title} {\enquote {\bibinfo {title}
  {Magnetization and universal sub-critical behaviour in two-dimensional {XY}
  magnets},}\ }\href {\doibase 10.1088/0953-8984/5/4/004} {\bibfield  {journal}
  {\bibinfo  {journal} {Journal of Physics: Condensed Matter}\ }\textbf
  {\bibinfo {volume} {5}},\ \bibinfo {pages} {L53--L59} (\bibinfo {year}
  {1993})}\BibitemShut {NoStop}%
\bibitem [{\citenamefont {Cuccoli}\ \emph {et~al.}(2003)\citenamefont
  {Cuccoli}, \citenamefont {Roscilde}, \citenamefont {Tognetti}, \citenamefont
  {Vaia},\ and\ \citenamefont {Verrucchi}}]{Cuccoli}%
  \BibitemOpen
  \bibfield  {author} {\bibinfo {author} {\bibfnamefont {A.}~\bibnamefont
  {Cuccoli}}, \bibinfo {author} {\bibfnamefont {T.}~\bibnamefont {Roscilde}},
  \bibinfo {author} {\bibfnamefont {V.}~\bibnamefont {Tognetti}}, \bibinfo
  {author} {\bibfnamefont {R.}~\bibnamefont {Vaia}}, \ and\ \bibinfo {author}
  {\bibfnamefont {P.}~\bibnamefont {Verrucchi}},\ }\bibfield  {title} {\enquote
  {\bibinfo {title} {Quantum monte carlo study of $s=\frac{1}{2}$ weakly
  anisotropic antiferromagnets on the square lattice},}\ }\href {\doibase
  10.1103/PhysRevB.67.104414} {\bibfield  {journal} {\bibinfo  {journal} {Phys.
  Rev. B}\ }\textbf {\bibinfo {volume} {67}},\ \bibinfo {pages} {104414}
  (\bibinfo {year} {2003})}\BibitemShut {NoStop}%
\bibitem [{\citenamefont {Rogado}\ \emph {et~al.}(2002)\citenamefont {Rogado},
  \citenamefont {Huang}, \citenamefont {Lyun}, \citenamefont {Ramirez},
  \citenamefont {Huse},\ and\ \citenamefont {Cava}}]{Rogado}%
  \BibitemOpen
  \bibfield  {author} {\bibinfo {author} {\bibfnamefont {N.}~\bibnamefont
  {Rogado}}, \bibinfo {author} {\bibfnamefont {Q.}~\bibnamefont {Huang}},
  \bibinfo {author} {\bibfnamefont {J.~W.}\ \bibnamefont {Lyun}}, \bibinfo
  {author} {\bibfnamefont {A.~P.}\ \bibnamefont {Ramirez}}, \bibinfo {author}
  {\bibfnamefont {D.}~\bibnamefont {Huse}}, \ and\ \bibinfo {author}
  {\bibfnamefont {R.~J.}\ \bibnamefont {Cava}},\ }\bibfield  {title} {\enquote
  {\bibinfo {title} {{B}a{N}i$_2${V}$_2${O}$_8$: A two-dimensional honeycomb
  antiferromagnet},}\ }\href {\doibase 10.1103/PhysRevB.65.144443} {\bibfield
  {journal} {\bibinfo  {journal} {Phys. Rev. B}\ }\textbf {\bibinfo {volume}
  {65}},\ \bibinfo {pages} {144443} (\bibinfo {year} {2002})}\BibitemShut
  {NoStop}%
\bibitem [{\citenamefont {Klyushina}\ \emph {et~al.}(2017)\citenamefont
  {Klyushina}, \citenamefont {Lake}, \citenamefont {Islam}, \citenamefont
  {Park}, \citenamefont {Schneidewind}, \citenamefont {Guidi}, \citenamefont
  {Goremychkin}, \citenamefont {Klemke},\ and\ \citenamefont
  {M\aa{}nsson}}]{KlyushinaBaNi}%
  \BibitemOpen
  \bibfield  {author} {\bibinfo {author} {\bibfnamefont {E.~S.}\ \bibnamefont
  {Klyushina}}, \bibinfo {author} {\bibfnamefont {B.}~\bibnamefont {Lake}},
  \bibinfo {author} {\bibfnamefont {A.~T. M.~N.}\ \bibnamefont {Islam}},
  \bibinfo {author} {\bibfnamefont {J.~T.}\ \bibnamefont {Park}}, \bibinfo
  {author} {\bibfnamefont {A.}~\bibnamefont {Schneidewind}}, \bibinfo {author}
  {\bibfnamefont {T.}~\bibnamefont {Guidi}}, \bibinfo {author} {\bibfnamefont
  {E.~A.}\ \bibnamefont {Goremychkin}}, \bibinfo {author} {\bibfnamefont
  {B.}~\bibnamefont {Klemke}}, \ and\ \bibinfo {author} {\bibfnamefont
  {M.}~\bibnamefont {M\aa{}nsson}},\ }\bibfield  {title} {\enquote {\bibinfo
  {title} {Investigation of the spin-1 honeycomb antiferromagnet
  {B}a{N}i$_2${V}$_2${O}$_8$ with easy-plane anisotropy},}\ }\href {\doibase
  10.1103/PhysRevB.96.214428} {\bibfield  {journal} {\bibinfo  {journal} {Phys.
  Rev. B}\ }\textbf {\bibinfo {volume} {96}},\ \bibinfo {pages} {214428}
  (\bibinfo {year} {2017})}\BibitemShut {NoStop}%
\bibitem [{\citenamefont {Sengupta}\ \emph {et~al.}(2003)\citenamefont
  {Sengupta}, \citenamefont {Sandvik},\ and\ \citenamefont
  {Singh}}]{Specificheat2D}%
  \BibitemOpen
  \bibfield  {author} {\bibinfo {author} {\bibfnamefont {Pinaki}\ \bibnamefont
  {Sengupta}}, \bibinfo {author} {\bibfnamefont {Anders~W.}\ \bibnamefont
  {Sandvik}}, \ and\ \bibinfo {author} {\bibfnamefont {Rajiv R.~P.}\
  \bibnamefont {Singh}},\ }\bibfield  {title} {\enquote {\bibinfo {title}
  {Specific heat of quasi-two-dimensional antiferromagnetic heisenberg models
  with varying interplanar couplings},}\ }\href {\doibase
  10.1103/PhysRevB.68.094423} {\bibfield  {journal} {\bibinfo  {journal} {Phys.
  Rev. B}\ }\textbf {\bibinfo {volume} {68}},\ \bibinfo {pages} {094423}
  (\bibinfo {year} {2003})}\BibitemShut {NoStop}%
\bibitem [{\citenamefont {Heinrich}\ \emph {et~al.}(2003)\citenamefont
  {Heinrich}, \citenamefont {Krug~von Nidda}, \citenamefont {Loidl},
  \citenamefont {Rogado},\ and\ \citenamefont {Cava}}]{Heinrich}%
  \BibitemOpen
  \bibfield  {author} {\bibinfo {author} {\bibfnamefont {M.}~\bibnamefont
  {Heinrich}}, \bibinfo {author} {\bibfnamefont {H.-A.}\ \bibnamefont {Krug~von
  Nidda}}, \bibinfo {author} {\bibfnamefont {A.}~\bibnamefont {Loidl}},
  \bibinfo {author} {\bibfnamefont {N.}~\bibnamefont {Rogado}}, \ and\ \bibinfo
  {author} {\bibfnamefont {R.~J.}\ \bibnamefont {Cava}},\ }\bibfield  {title}
  {\enquote {\bibinfo {title} {Potential signature of a {K}osterlitz-{T}houless
  transition in {B}a{N}i$_2${V}$_2${O}$_8$},}\ }\href {\doibase
  10.1103/PhysRevLett.91.137601} {\bibfield  {journal} {\bibinfo  {journal}
  {Phys. Rev. Lett.}\ }\textbf {\bibinfo {volume} {91}},\ \bibinfo {pages}
  {137601} (\bibinfo {year} {2003})}\BibitemShut {NoStop}%
\bibitem [{\citenamefont {Waibel}\ \emph {et~al.}(2015)\citenamefont {Waibel},
  \citenamefont {Fischer}, \citenamefont {Wolf}, \citenamefont {L\"ohneysen},\
  and\ \citenamefont {Pilawa}}]{Waibel2015}%
  \BibitemOpen
  \bibfield  {author} {\bibinfo {author} {\bibfnamefont {D.}~\bibnamefont
  {Waibel}}, \bibinfo {author} {\bibfnamefont {G.}~\bibnamefont {Fischer}},
  \bibinfo {author} {\bibfnamefont {Th.}\ \bibnamefont {Wolf}}, \bibinfo
  {author} {\bibfnamefont {H.~v.}\ \bibnamefont {L\"ohneysen}}, \ and\ \bibinfo
  {author} {\bibfnamefont {B.}~\bibnamefont {Pilawa}},\ }\bibfield  {title}
  {\enquote {\bibinfo {title} {Determining the
  {B}erezinskii-{K}osterlitz-{T}houless coherence length in
  {B}a{N}i$_2${V}$_2${O}$_8$ by $^{51}\mathrm{V}$ nmr},}\ }\href {\doibase
  10.1103/PhysRevB.91.214412} {\bibfield  {journal} {\bibinfo  {journal} {Phys.
  Rev. B}\ }\textbf {\bibinfo {volume} {91}},\ \bibinfo {pages} {214412}
  (\bibinfo {year} {2015})}\BibitemShut {NoStop}%
\bibitem [{\citenamefont {Semadeni}\ \emph {et~al.}(2001)\citenamefont
  {Semadeni}, \citenamefont {Roessli},\ and\ \citenamefont {B{\"o}ni}}]{TASP}%
  \BibitemOpen
  \bibfield  {author} {\bibinfo {author} {\bibfnamefont {F.}~\bibnamefont
  {Semadeni}}, \bibinfo {author} {\bibfnamefont {B.}~\bibnamefont {Roessli}}, \
  and\ \bibinfo {author} {\bibfnamefont {P.}~\bibnamefont {B{\"o}ni}},\
  }\bibfield  {title} {\enquote {\bibinfo {title} {Three-axis spectroscopy with
  remanent benders},}\ }\href {\doibase
  https://doi.org/10.1016/S0921-4526(00)00860-7} {\bibfield  {journal}
  {\bibinfo  {journal} {Physica B Condens. Matter}\ }\textbf {\bibinfo {volume}
  {297}},\ \bibinfo {pages} {152 -- 154} (\bibinfo {year} {2001})}\BibitemShut
  {NoStop}%
\bibitem [{\citenamefont {Taroni}\ \emph {et~al.}(2008)\citenamefont {Taroni},
  \citenamefont {Bramwell},\ and\ \citenamefont {Holdsworth}}]{Taroni}%
  \BibitemOpen
  \bibfield  {author} {\bibinfo {author} {\bibfnamefont {A}~\bibnamefont
  {Taroni}}, \bibinfo {author} {\bibfnamefont {S~T}\ \bibnamefont {Bramwell}},
  \ and\ \bibinfo {author} {\bibfnamefont {P~C~W}\ \bibnamefont {Holdsworth}},\
  }\bibfield  {title} {\enquote {\bibinfo {title} {Universal window for
  two-dimensional critical exponents},}\ }\href {\doibase
  10.1088/0953-8984/20/27/275233} {\bibfield  {journal} {\bibinfo  {journal}
  {J. Phys. Condens. Matter}\ }\textbf {\bibinfo {volume} {20}},\ \bibinfo
  {pages} {275233} (\bibinfo {year} {2008})}\BibitemShut {NoStop}%
\bibitem [{\citenamefont {Collins}()}]{collins}%
  \BibitemOpen
  \bibfield  {author} {\bibinfo {author} {\bibfnamefont {M.~F.}\ \bibnamefont
  {Collins}},\ }\href@noop {} {\enquote {\bibinfo {title} {Magnetic critical
  scattering},}\ }\bibinfo {note} {(Oxford University Press, Great Britain
  (1989))}\BibitemShut {NoStop}%
\bibitem [{\citenamefont {Elstner}\ \emph {et~al.}(1995)\citenamefont
  {Elstner}, \citenamefont {Sokol}, \citenamefont {Singh}, \citenamefont
  {Greven},\ and\ \citenamefont {Birgeneau}}]{2DHcor}%
  \BibitemOpen
  \bibfield  {author} {\bibinfo {author} {\bibfnamefont {N.}~\bibnamefont
  {Elstner}}, \bibinfo {author} {\bibfnamefont {A.}~\bibnamefont {Sokol}},
  \bibinfo {author} {\bibfnamefont {R.~R.~P.}\ \bibnamefont {Singh}}, \bibinfo
  {author} {\bibfnamefont {M.}~\bibnamefont {Greven}}, \ and\ \bibinfo {author}
  {\bibfnamefont {R.~J.}\ \bibnamefont {Birgeneau}},\ }\bibfield  {title}
  {\enquote {\bibinfo {title} {Spin dependence of correlations in
  two-dimensional square-lattice quantum heisenberg antiferromagnets},}\ }\href
  {\doibase 10.1103/PhysRevLett.75.938} {\bibfield  {journal} {\bibinfo
  {journal} {Phys. Rev. Lett.}\ }\textbf {\bibinfo {volume} {75}},\ \bibinfo
  {pages} {938--941} (\bibinfo {year} {1995})}\BibitemShut {NoStop}%
\bibitem [{\citenamefont {Jos\'e}\ \emph {et~al.}(1977)\citenamefont {Jos\'e},
  \citenamefont {Kadanoff}, \citenamefont {Kirkpatrick},\ and\ \citenamefont
  {Nelson}}]{Jose}%
  \BibitemOpen
  \bibfield  {author} {\bibinfo {author} {\bibfnamefont {Jorge~V.}\
  \bibnamefont {Jos\'e}}, \bibinfo {author} {\bibfnamefont {Leo~P.}\
  \bibnamefont {Kadanoff}}, \bibinfo {author} {\bibfnamefont {Scott}\
  \bibnamefont {Kirkpatrick}}, \ and\ \bibinfo {author} {\bibfnamefont
  {David~R.}\ \bibnamefont {Nelson}},\ }\bibfield  {title} {\enquote {\bibinfo
  {title} {Renormalization, vortices, and symmetry-breaking perturbations in
  the two-dimensional planar model},}\ }\href {\doibase
  10.1103/PhysRevB.16.1217} {\bibfield  {journal} {\bibinfo  {journal} {Phys.
  Rev. B}\ }\textbf {\bibinfo {volume} {16}},\ \bibinfo {pages} {1217--1241}
  (\bibinfo {year} {1977})}\BibitemShut {NoStop}%
\bibitem [{\citenamefont {Klyushina}\ \emph {et~al.}()\citenamefont
  {Klyushina}, \citenamefont {Lake},\ and\ \citenamefont {Lord}}]{MuonData}%
  \BibitemOpen
  \bibfield  {author} {\bibinfo {author} {\bibfnamefont {E.}~\bibnamefont
  {Klyushina}}, \bibinfo {author} {\bibfnamefont {B.}~\bibnamefont {Lake}}, \
  and\ \bibinfo {author} {\bibfnamefont {J.~S.}\ \bibnamefont {Lord}},\
  }\bibfield  {title} {\enquote {\bibinfo {title} {Dynamics of the quasi two
  dimensional xxz honeycomb antiferromagnet bani2v2o8},}\ }\href {\doibase
  10.5286/ISIS.E.RB1820523} {\bibfield  {journal} {\bibinfo  {journal} {STFC
  ISIS Facility}\ }\textbf {\bibinfo {volume} {RB1820523}},\
  10.5286/ISIS.E.RB1820523}\BibitemShut {NoStop}%
\bibitem [{\citenamefont {Frandsen}\ \emph {et~al.}(2016)\citenamefont
  {Frandsen}, \citenamefont {Liu}, \citenamefont {Cheung},\ and\ \citenamefont
  {al}}]{TFfunction}%
  \BibitemOpen
  \bibfield  {author} {\bibinfo {author} {\bibfnamefont {B.}~\bibnamefont
  {Frandsen}}, \bibinfo {author} {\bibfnamefont {L.}~\bibnamefont {Liu}},
  \bibinfo {author} {\bibfnamefont {S.}~\bibnamefont {Cheung}}, \ and\ \bibinfo
  {author} {\bibfnamefont {et}~\bibnamefont {al}},\ }\bibfield  {title}
  {\enquote {\bibinfo {title} {Volume-wise destruction of the antiferromagnetic
  mott insulating state through quantum tuning},}\ }\href {\doibase
  10.1038/ncomms12519} {\bibfield  {journal} {\bibinfo  {journal} {Nat Commun}\
  }\textbf {\bibinfo {volume} {7}},\ \bibinfo {pages} {12519} (\bibinfo {year}
  {2016})}\BibitemShut {NoStop}%
\bibitem [{\citenamefont {Bendele}\ \emph {et~al.}(2010)\citenamefont
  {Bendele}, \citenamefont {Babkevich}, \citenamefont {Katrych}, \citenamefont
  {Gvasaliya}, \citenamefont {Pomjakushina}, \citenamefont {Conder},
  \citenamefont {Roessli}, \citenamefont {Boothroyd}, \citenamefont
  {Khasanov},\ and\ \citenamefont {Keller}}]{PhysRevB.82.212504}%
  \BibitemOpen
  \bibfield  {author} {\bibinfo {author} {\bibfnamefont {M.}~\bibnamefont
  {Bendele}}, \bibinfo {author} {\bibfnamefont {P.}~\bibnamefont {Babkevich}},
  \bibinfo {author} {\bibfnamefont {S.}~\bibnamefont {Katrych}}, \bibinfo
  {author} {\bibfnamefont {S.~N.}\ \bibnamefont {Gvasaliya}}, \bibinfo {author}
  {\bibfnamefont {E.}~\bibnamefont {Pomjakushina}}, \bibinfo {author}
  {\bibfnamefont {K.}~\bibnamefont {Conder}}, \bibinfo {author} {\bibfnamefont
  {B.}~\bibnamefont {Roessli}}, \bibinfo {author} {\bibfnamefont {A.~T.}\
  \bibnamefont {Boothroyd}}, \bibinfo {author} {\bibfnamefont {R.}~\bibnamefont
  {Khasanov}}, \ and\ \bibinfo {author} {\bibfnamefont {H.}~\bibnamefont
  {Keller}},\ }\bibfield  {title} {\enquote {\bibinfo {title} {Tuning the
  superconducting and magnetic properties of {F}e$_y${S}e$_{0.25}${T}e$_{0.75}$
  by varying the iron content},}\ }\href {\doibase 10.1103/PhysRevB.82.212504}
  {\bibfield  {journal} {\bibinfo  {journal} {Phys. Rev. B}\ }\textbf {\bibinfo
  {volume} {82}},\ \bibinfo {pages} {212504} (\bibinfo {year}
  {2010})}\BibitemShut {NoStop}%
\bibitem [{\citenamefont {Khasanov}\ \emph {et~al.}(2008)\citenamefont
  {Khasanov}, \citenamefont {Shengelaya}, \citenamefont {Di~Castro},
  \citenamefont {Morenzoni}, \citenamefont {Maisuradze}, \citenamefont
  {Savi\ifmmode~\acute{c}\else \'{c}\fi{}}, \citenamefont {Conder},
  \citenamefont {Pomjakushina}, \citenamefont {Bussmann-Holder},\ and\
  \citenamefont {Keller}}]{PhysRevLett.101.077001}%
  \BibitemOpen
  \bibfield  {author} {\bibinfo {author} {\bibfnamefont {R.}~\bibnamefont
  {Khasanov}}, \bibinfo {author} {\bibfnamefont {A.}~\bibnamefont
  {Shengelaya}}, \bibinfo {author} {\bibfnamefont {D.}~\bibnamefont
  {Di~Castro}}, \bibinfo {author} {\bibfnamefont {E.}~\bibnamefont
  {Morenzoni}}, \bibinfo {author} {\bibfnamefont {A.}~\bibnamefont
  {Maisuradze}}, \bibinfo {author} {\bibfnamefont {I.~M.}\ \bibnamefont
  {Savi\ifmmode~\acute{c}\else \'{c}\fi{}}}, \bibinfo {author} {\bibfnamefont
  {K.}~\bibnamefont {Conder}}, \bibinfo {author} {\bibfnamefont
  {E.}~\bibnamefont {Pomjakushina}}, \bibinfo {author} {\bibfnamefont
  {A.}~\bibnamefont {Bussmann-Holder}}, \ and\ \bibinfo {author} {\bibfnamefont
  {H.}~\bibnamefont {Keller}},\ }\bibfield  {title} {\enquote {\bibinfo {title}
  {Oxygen isotope effects on the superconducting transition and magnetic states
  within the phase diagram of
  {Y}$_{1\ensuremath{-}x}${P}r$_{x}${B}a$_{2}${C}u$_{3}${O}$_{7\ensuremath{-}\ensuremath{\delta}}$},}\
  }\href {\doibase 10.1103/PhysRevLett.101.077001} {\bibfield  {journal}
  {\bibinfo  {journal} {Phys. Rev. Lett.}\ }\textbf {\bibinfo {volume} {101}},\
  \bibinfo {pages} {077001} (\bibinfo {year} {2008})}\BibitemShut {NoStop}%
\bibitem [{\citenamefont {Sandvik}(1999)}]{PhysRevB.59.R14157}%
  \BibitemOpen
  \bibfield  {author} {\bibinfo {author} {\bibfnamefont {Anders~W.}\
  \bibnamefont {Sandvik}},\ }\bibfield  {title} {\enquote {\bibinfo {title}
  {Stochastic series expansion method with operator-loop update},}\ }\href
  {\doibase 10.1103/PhysRevB.59.R14157} {\bibfield  {journal} {\bibinfo
  {journal} {Phys. Rev. B}\ }\textbf {\bibinfo {volume} {59}},\ \bibinfo
  {pages} {R14157--R14160} (\bibinfo {year} {1999})}\BibitemShut {NoStop}%
\bibitem [{\citenamefont {Sandvik}\ and\ \citenamefont
  {Kurkij\"arvi}(1991)}]{PhysRevB.43.5950}%
  \BibitemOpen
  \bibfield  {author} {\bibinfo {author} {\bibfnamefont {Anders~W.}\
  \bibnamefont {Sandvik}}\ and\ \bibinfo {author} {\bibfnamefont {Juhani}\
  \bibnamefont {Kurkij\"arvi}},\ }\bibfield  {title} {\enquote {\bibinfo
  {title} {Quantum monte carlo simulation method for spin systems},}\ }\href
  {\doibase 10.1103/PhysRevB.43.5950} {\bibfield  {journal} {\bibinfo
  {journal} {Phys. Rev. B}\ }\textbf {\bibinfo {volume} {43}},\ \bibinfo
  {pages} {5950--5961} (\bibinfo {year} {1991})}\BibitemShut {NoStop}%
\bibitem [{\citenamefont {Henelius}\ and\ \citenamefont
  {Sandvik}(2000)}]{PhysRevB.62.1102}%
  \BibitemOpen
  \bibfield  {author} {\bibinfo {author} {\bibfnamefont {Patrik}\ \bibnamefont
  {Henelius}}\ and\ \bibinfo {author} {\bibfnamefont {Anders~W.}\ \bibnamefont
  {Sandvik}},\ }\bibfield  {title} {\enquote {\bibinfo {title} {Sign problem in
  monte carlo simulations of frustrated quantum spin systems},}\ }\href
  {\doibase 10.1103/PhysRevB.62.1102} {\bibfield  {journal} {\bibinfo
  {journal} {Phys. Rev. B}\ }\textbf {\bibinfo {volume} {62}},\ \bibinfo
  {pages} {1102--1113} (\bibinfo {year} {2000})}\BibitemShut {NoStop}%
\bibitem [{\citenamefont {Toth}\ and\ \citenamefont {Lake}(2015)}]{SpinW}%
  \BibitemOpen
  \bibfield  {author} {\bibinfo {author} {\bibfnamefont {S}~\bibnamefont
  {Toth}}\ and\ \bibinfo {author} {\bibfnamefont {B}~\bibnamefont {Lake}},\
  }\bibfield  {title} {\enquote {\bibinfo {title} {Linear spin wave theory for
  single-q incommensurate magnetic structures},}\ }\href
  {http://stacks.iop.org/0953-8984/27/i=16/a=166002} {\bibfield  {journal}
  {\bibinfo  {journal} {J. Phys. Condens. Matter}\ }\textbf {\bibinfo {volume}
  {27}},\ \bibinfo {pages} {166002} (\bibinfo {year} {2015})}\BibitemShut
  {NoStop}%
\end{thebibliography}
%

\end{document}